\documentclass[a4paper,11pt]{article}
\pdfoutput=1 % if your are submitting a pdflatex (i.e. if you have
\usepackage{jheppub} % for details on the use of the package, please
\usepackage[T1]{fontenc} % if needed
\usepackage[utf8]{inputenc}
\usepackage{amsmath,amssymb,amstext,amsfonts} % AMS' mathematical symbols

\usepackage[dvipsnames,table]{xcolor}
\usepackage{graphicx}
\usepackage{hyperref}
\usepackage{cancel} %for strike-through font
\usepackage{float}
\usepackage{bbm}
\usepackage{mathtools}
\usepackage{lscape}
\usepackage{siunitx}
\usepackage{subcaption}
\usepackage{cleveref}
\usepackage{amsmath}
\usepackage{rotating}
\usepackage{adjustbox}
\usepackage{comment}
\usepackage{tabularx}

\usepackage{slashed}
\usepackage{physics}
\usepackage{nicematrix}
\makeatletter
\renewcommand*\env@matrix[1][\arraystretch]{%
  \edef\arraystretch{#1}%
  \hskip -\arraycolsep
  \let\@ifnextchar\new@ifnextchar
  \array{*\c@MaxMatrixCols c}}
\makeatother

\makeatletter
\g@addto@macro\bfseries{\boldmath}
\makeatother

\definecolor{mygray1}{gray}{0.4}

\newcommand{\diff}{\mathrm{d}}

\def\beq{\begin{equation}}
\def\eeq{\end{equation}}
\def\bsp#1\esp{\begin{split}#1\end{split}}

\usepackage{tikz}
\usepackage{bm}
    \usetikzlibrary{positioning}
    \usetikzlibrary{arrows}
    \usetikzlibrary{shapes}
    \usepackage{pgfplots}
    \usetikzlibrary{calc}
    \usetikzlibrary{decorations.markings}
     \usetikzlibrary{decorations.pathmorphing}
    \tikzset{snake it/.style={decorate, decoration=snake}}
\def\centerarc[#1](#2)(#3:#4:#5) % Syntax: [draw options] (center) (initial angle:final angle:radius)
    { \draw[#1] ($(#2)+({#5*cos(#3)},{#5*sin(#3)})$) arc (#3:#4:#5); }
    \usepackage{booktabs}
\usepackage[compat=1.1.0]{tikz-feynman}
\usepackage{contour}

\title{\boldmath 
All three-loop four-point massless integrals with one off-shell leg}

\author[a]{Thomas Gehrmann,}
\author[b]{Petr Jakub\v{c}\'{i}k,}
\author[c,d]{Jungwon Lim,}
\author[e]{Andreas von Manteuffel,}
\author[b]{Cesare Carlo Mella,}
\author[f]{Lorenzo Tancredi}

\affiliation[a]{Physik-Institut, Universit\"at Zurich, Winterthurerstrasse 190, CH-8057 Z\"urich, Switzerland}
\affiliation[b]{Rudolf Peierls Centre for Theoretical Physics, University of Oxford,
Clarendon Laboratory, Parks Road, Oxford OX1 3PU}
\affiliation[c]{Deutsches Elektronen-Synchrotron DESY,
Notkestr. 85, 22607 Hamburg, Germany}
\affiliation[d]{Max-Planck-Institut f\"ur Physik, Werner-Heisenberg-Institut, Boltzmannstr. 8, 85748 Garching, Germany}
\affiliation[e]{
Institut f\"ur Theoretische Physik, Universität Regensburg, 93040 Regensburg, Germany}
\affiliation[f]{Technical University of Munich, TUM School of Natural Sciences, Physics Department, James-Franck-Straße 1, D-85748 Garching, Germany}

\emailAdd{thomas.gehrmann@uzh.ch}
\emailAdd{petr.jakubcik@physics.ox.ac.uk}
\emailAdd{wonlim@desy.de}
\emailAdd{manteuffel@ur.de}
\emailAdd{cesare.mella@physics.ox.ac.uk}
\emailAdd{lorenzo.tancredi@tum.de}

\preprint{\begin{minipage}[t]{8cm}\begin{flushright} 
OUTP-26-06P \\ TUM-HEP-1615/26 \\ ZU-TH 39/26 \\ MPP-2026-190 
      \end{flushright}\end{minipage}}

\usepackage{xspace}

\usetikzlibrary{arrows.meta,positioning,fit,backgrounds}
 
\definecolor{colA}{HTML}{1D9E75}
\definecolor{colB}{HTML}{D85A30}
\definecolor{colC}{HTML}{7F77DD}

\tikzset{
  blob/.style={
    draw, rounded corners=1mm, align=center, line width=0.4pt,
    minimum width=42mm, minimum height=12mm, inner sep=1.2mm,
    font=\fontsize{10}{11.5}\selectfont},
  stepA/.style={blob, draw=colA!75!black, fill=colA!7},
  stepB/.style={blob, draw=colB!75!black, fill=colB!7},
  stepC/.style={blob, draw=colC!75!black, fill=colC!7},
  flow/.style={-{Stealth[length=1.7mm,width=1.3mm]}, line width=0.45pt, draw=black!72},
  feed/.style={flow, densely dashed, draw=black!55},
  feeder/.style={line width=0.45pt, draw=black!72},
  dec/.style={font=\fontsize{8}{9}\selectfont\itshape, align=center,
              inner sep=0.5mm, fill=white, fill opacity=0.95, text opacity=1},
  region/.style={draw, densely dashed, rounded corners=2mm, line width=0.4pt,
                 inner xsep=2mm, inner ysep=3.5mm},
  rlabel/.style={font=\fontsize{9.5}{11}\selectfont\bfseries,
                 anchor=south west, inner sep=0.6mm}
}

\abstract{
We present the complete set of massless three-loop four-point master 
integrals with one off-shell external leg.
These are essential building blocks for the three-loop virtual corrections to Higgs-boson or vector-boson production in association with a jet at the Large Hadron Collider, required for next-to-next-to-next-to-leading-order QCD predictions.
We~construct a complete basis of canonical integrals
 by identifying integrands with simple poles and unit leading singularities, solely by integrand analysis.
This includes an integral family with 371 master integrals and a 19-dimensional top sector.
The solutions for all relevant integrals are provided as a map to a minimal set of iterated integrals graded according to their analytic properties, together with high-precision series expansions. 

}

\begin{document} 
\maketitle
\flushbottom

\section{Introduction}
\label{sec:intro}
Precision predictions for $pp\to H+j$ and $pp\to V+j$ with
$V=Z,W^\pm,\gamma^*$ are central to the present and future physics programme
of the Large Hadron Collider (LHC).  Higgs-plus-jet production gives direct access to the transverse
momentum of the Higgs boson and to the boosted regime, where differential
distributions provide stringent tests of the Higgs--gluon interaction and
enhanced sensitivity to heavy virtual particles.  Vector-boson-plus-jet
production, with its large rates and clean leptonic signatures, is a
precision Standard-Model benchmark, a probe of parton distributions and the
strong coupling, and an important background to Higgs and missing-energy
searches.  As the experimental uncertainties shrink towards the HL-LHC era,
theoretical control of these processes must improve accordingly~\cite{Cepeda:2019klc}.  Fully differential QCD predictions are presently known
through next-to-next-to-leading order (NNLO) for Higgs-plus-jet
\cite{Boughezal:2015dra,Boughezal:2015aha,Caola:2015wna,Chen:2016zka} and for
$W$- and $Z$-boson production in association with a jet
\cite{Boughezal:2015dva,Boughezal:2015ded,Gehrmann-DeRidder:2015wbt,Gehrmann-DeRidder:2017mvr,Alioli:2025hpa,NNLOJET:2025rno}.  Extending these predictions to
N$^3$LO requires, among several real and mixed real--virtual contributions, the
purely virtual interference of the three-loop $2\to2$ amplitude with the tree
amplitude.  The scalar building blocks of this contribution are precisely the
massless three-loop four-point integrals with one off-shell external leg studied
in this paper.

Substantial progress has recently been made at the amplitude level.  The three-loop helicity amplitudes for the production of a vector boson and a jet
are known in the planar approximation
\cite{Gehrmann:2023jyv}.  For Higgs-plus-jet production in the heavy-top
effective theory, the interference with the tree amplitude was first obtained
at generalised leading colour~\cite{Chen:2025utl}, and subsequently extended to helicity amplitudes in full colour in~\cite{Guan:2026igf}.
The latter result also shows that subleading-colour
terms need not be numerically suppressed. 
The supersymmetric counterpart of this amplitude, the three-point form factor of the operator $\Tr(\phi^2)$, was evaluated using loop integrals first at leading colour in~\cite{Gehrmann:2024tds} and at full colour in~\cite{Guan:2025awp}.

The Feynman integrals 
that appear in these amplitudes can be classified into
integral families and 
reduced to 
linear combinations of master integrals by employing 
integration-by-parts (IBP) identities~\cite{Chetyrkin:1981qh,Laporta:2000dsw}.
The evaluation of the required 
master integrals is commonly achieved by solving differential equations in kinematic invariants for the integrals~\cite{Kotikov:1991pm, Remiddi:1997ny, Gehrmann:1999as}.
The two-loop master
integrals were first computed using this method in \cite{Gehrmann:2000zt,Gehrmann:2001ck}, 
subsequently employed in 
the two-loop helicity amplitudes for 
$V$+jet~\cite{Garland:2002ak,Gehrmann:2011ab} and $H$+jet~\cite{Gehrmann:2011aa},
and extended to higher orders in the dimensional regulator $\epsilon$ in~\cite{Gehrmann:2023etk}.
At three
loops the planar families have been evaluated
in refs.~\cite{DiVita:2014pza,Canko:2021xmn,Syrrakos:2023mor}, with two non-planar families added in~\cite{Henn:2023vbd}.
All integral topologies required for $H$+jet production at leading colour were
published in ref.~\cite{Gehrmann:2024tds}. 
This work also detailed the analytic structure and function space of the resulting amplitude, confirming hypotheses from cluster algebras on the adjacency of the letters entering its iterated-integral representation~\cite{Dixon:2020bbt,Chicherin:2020umh,Aliaj:2024zgp}.
The calculation of the full-colour amplitude in ref.~\cite{Guan:2026igf} required the evaluation of the remaining integral families, but the canonical bases and solutions for the individual master integrals were not published there.

In this paper, we provide the complete canonical bases and integral solutions, adding nine non-planar families to the published results.
These are of interest in their own right, both for reuse in other amplitudes and for investigations of analytic structure and function spaces.
Central to this calculation is the choice of integral basis: we seek one in which the differential equations take the \emph{canonical} form~\cite{Henn:2013pwa} with factorised kinematic and $\epsilon$ dependence.
This form makes the relevant iterated-integral function space manifest to all orders in $\epsilon$ and allows the equations to be solved in terms of iterated integrals~\cite{Chen:1977oja} with well-understood identities.

There are two complementary approaches to constructing such a basis.  The first starts from a generic set of master integrals obtained after reduction via linear identities, and searches for a kinematic- and $\epsilon$-dependent change of basis that transforms their differential equations into the canonical form.
The method is automated in public implementations including
\texttt{Fuchsia}, \texttt{epsilon}, \texttt{CANONICA}, and
\texttt{Libra}~\cite{Lee:2014ioa,Gituliar:2017vzm,Prausa:2017ltv,Meyer:2017joq,Lee:2020zfb}.  This is the strategy adopted in ref.~\cite{Guan:2026igf} to compute the $H$+jet amplitude, where the Lee algorithm \cite{Lee:2014ioa}  was applied on a univariate kinematic slice, followed by a multivariate transformation based on
the approach of \texttt{CANONICA} \cite{Meyer:2017joq}, see also~\cite{Gehrmann:2014bfa}.

The second approach, which we pursue here, constructs integrands with only simple poles and unit leading singularities, meaning that their non-zero maximal iterated residues are $\pm1$.
By identifying candidate canonical integrals directly from their integrands, it avoids the reconstruction of intermediate non-canonical differential equations.
This method is motivated by early studies of planar ${\cal N}=4$ super-Yang--Mills amplitudes, which showed that multi-loop integrands could be organised in terms of local chiral integrals normalised to have unit leading singularities~\cite{Arkani-Hamed:2010zjl,Arkani-Hamed:2010pyv}.  
This viewpoint was sharpened in~\cite{Bern:2014kca}, where the supersymmetric three-loop four-point amplitude was constructed from a complete basis of integrands with only logarithmic singularities and no poles at infinity. Ref.~\cite{Henn:2013pwa} connected these observations to the differential equation method by proposing that integrals of uniform transcendental weight naturally satisfy differential equations in the canonical $\epsilon$-factorised form, and advocated kinematics-independent leading singularities as a practical criterion for identifying such integrals.  
The connection can also be seen directly at the integrand level: for integrals admitting suitable d-log representations, simple differential equations can be derived using their singularity structure and generalised unitarity~\cite{Herrmann:2019upk}.  
The implication from constant leading singularities to uniform transcendentality and a canonical differential equation is thus a powerful conjectural criterion for the construction of a candidate basis.

These ideas have also been implemented in code, most notably in \texttt{DlogBasis}~\cite{Henn:2020lye}, which searches for numerator choices yielding integrands with single poles and constant leading singularities. 
In this paper, we present an improved search strategy for 
integrands with constant leading singularities, 
which involves extended classes of variable changes and retains non-rationalised square roots at intermediate 
steps of the residue calculation. These improvements 
are crucial to overcome algebraic obstructions that 
prevented the canonical basis construction for 
the remaining non-planar families up to now. 
With the canonical bases at hand, we extend the methods developed for the leading-colour three-loop integrals in ref.~\cite{Gehrmann:2024tds} to solve the full set of master integrals.
These methods construct a minimal set of transcendental functions graded by their analytic properties and fix boundary conditions without the need for a full two-dimensional functional representation.
Furthermore, we find a high-precision series expansion in the decay kinematic region, enabling systematic analytic continuation to scattering regions.

The paper is organised as follows.  In \cref{sec:notation} we define the
kinematics and integral families and summarise the reduction to master integrals and
differential-equation setup.  The construction of integrands with unit leading
singularities is developed in \cref{sec:dlog-strategy}, including the treatment
of algebraic branches generated during the residue calculation.  We apply the
method to the most complicated integral family in \cref{sec:hfamily}, and present the complete
canonical basis and its solution in terms of iterated integrals in \cref{sec:results}.
Finally, we discuss two complementary strategies for their numerical evaluation in~\cref{sec:numerics}, before concluding in
\cref{sec:concl}.

\section{Computational setup}
\label{sec:notation}

\subsection{Kinematics}
We consider the decay of an off-shell particle of momentum $p_4$ into three
massless ones,
\begin{equation}
  p_4\to p_1+p_2+p_3\,,
  \qquad p_4=p_1+p_2+p_3\,.
  \label{eq:decay-kinematics}
\end{equation}
The on-shell conditions are
\begin{equation}
p_i^2 = 0\quad(i=1,2,3)\,, \qquad p_4^2 = m^2\,.
\end{equation}
We define the Mandelstam invariants in the usual way as
\begin{equation}
  s=(p_1+p_2)^2\,,\qquad
  t=(p_1+p_3)^2\,,\qquad
  u=(p_2+p_3)^2\,,
  \label{eq:mandelstam-invariants}
\end{equation}
so that
\begin{equation}
  s+t+u=m^2\,.
  \label{eq:stu-relation}
\end{equation}
It is often convenient to remove the overall scale and work with
\begin{equation}
  x=\frac{s}{m^2}\,,\qquad
  y=\frac{t}{m^2}\,,\qquad
  z=\frac{u}{m^2}\,,
  \label{eq:dimensionless-kinematics}
\end{equation}
and \cref{eq:stu-relation} becomes
\begin{equation}
  x+y+z=1\,.
  \label{eq:xyz-relation}
\end{equation}
For timelike $p_4$, with $m^2>0$, the decay region is identified by $x,y,z>0$.
Other physical channels are obtained by crossing and analytic continuation~\cite{Gehrmann:2002zr}.
To fully describe our result we will also need a square root $r$; its square is proportional to the Gram determinant of the three massless momenta
\begin{equation}
  r^2=m^2st\,(s+t-m^2)=-m^2stu\,.
  \label{eq:kinematic-gram-determinant}
\end{equation}
Branch choices for this root and the continuation to scattering regions are discussed in \cref{sec:analytic-continuation}.

\subsection{Integral families}
Following the standard approach to multiloop calculations, 
we organise the integrals appearing in our problem by the topology of the underlying graph. 
At three loops, the relevant graphs have at most ten massless internal propagators
and, ignoring insertions of two-point subgraphs, are covered by the $46$ families
shown in \cref{fig:one-mass-integral-families}, labelled $A$--$S_4$.
The numerical suffixes distinguish inequivalent assignments of the external mass to the same underlying massless graph, in the notation of the nine four-point all-massless irreducible families $A$--$H$ used in \cite{Henn:2020lye}.

We promote each ten-propagator graph to a complete \emph{integral
family}. With three loop momenta $k_1,k_2,k_3$ and three independent
external momenta $p_1,p_2,p_3$, there are $15$ independent scalar products involving at least one loop momentum.
The defining graph supplies ten inverse propagators; five auxiliary inverse propagators complete a basis for these scalar products.
The ordered list $Q_i=q_i^2+i0$, with $q_i$ a linear combination of loop and external momenta, defines the integral family $\mathcal F$.
The first ten entries correspond to the graph propagators and the last five are auxiliary.
We often use the shorthand $D_i=q_i^2$, suppressing the Feynman prescription.
The ordered momentum sets for all families are supplied in the electronic ancillary file \texttt{integralfamilies.yaml}.

Working in dimensional regularisation with the number of dimensions $d=4-2\epsilon$, any scalar integral in the
family can be written as
\begin{equation}
 I^{\mathcal F}_{\boldsymbol a}(s,t,m^2;\epsilon)
 =e^{3\gamma_E\epsilon}(-m^2)^{3\epsilon}
 \int\prod_{\ell=1}^{3}\frac{\diff^d k_\ell}{i\pi^{d/2}}
 \prod_{j=1}^{15}\bigl(Q_j^{\mathcal F}\bigr)^{-a_j}\,,
 \qquad \boldsymbol a=(a_1,\ldots,a_{15})\in\mathbb Z^{15}\,,
 \label{eq:scalar-integral-definition}
\end{equation}
where $\gamma_E$ is the Euler-Mascheroni constant.
The branch of the scale factor is understood with $-m^2-i0$.

Positive $a_j$ denote denominator powers and vanishing or negative
$a_j$ correspond to absent propagators or numerator insertions.
For the integrals considered here, the last five indices are non-positive, so auxiliary inverse propagators occur only in numerators.
A \emph{sector} specifies which graph propagators are present:
$S=\{j\mid a_j>0\}$, and a compact label for it is the sector number
\begin{equation}
  \sigma(S)=\sum_{j\in S}2^{j-1}\,.
  \label{eq:sector-number}
\end{equation}
Consequently, the top sector containing all ten graph propagators has sector number 1023.
$S'$ is called a \emph{sub-sector} of $S$ if $S'\subset S$ (and
equivalently $S$ is a \emph{super-sector} of $S'$).
Three integers associated with the vector
$\boldsymbol a$ will be useful bookkeeping devices throughout: the number
of propagators $t=|S|$, the total power of denominators
$r=\sum_{j\in S}a_j$, and the total power of numerators
$s=-\sum_{j\notin S}a_j$, called \emph{rank} of the integral.

\begin{figure}[p!]
  \centering
  \setlength{\tabcolsep}{0.55mm}
  \renewcommand{\arraystretch}{0}
  \newcommand{\onemassirreducibleplot}[1]{%
    \includegraphics[width=0.162\textwidth,height=0.114\textheight,keepaspectratio]{%
      #1.pdf}%
  }
  \newcommand{\onemassreducibleplot}[1]{%
    \includegraphics[width=0.162\textwidth,height=0.114\textheight,keepaspectratio]{%
      #1.pdf}%
  }

  \begin{tabular}{@{}*{6}{c}@{}}
    \onemassirreducibleplot{A}  & \onemassirreducibleplot{B1} & \onemassirreducibleplot{B2} &
    \onemassirreducibleplot{C}  & \onemassirreducibleplot{D}  & \onemassirreducibleplot{E1} \\[1.2mm]
    \onemassirreducibleplot{E2} & \onemassirreducibleplot{F1} & \onemassirreducibleplot{F2} &
    \onemassirreducibleplot{G1} & \onemassirreducibleplot{G2} & \onemassirreducibleplot{G3} \\[1.2mm]
    \onemassirreducibleplot{G4} & \onemassirreducibleplot{H}  & \onemassirreducibleplot{I1} &
    \onemassirreducibleplot{I2} & \onemassirreducibleplot{I3} & \onemassirreducibleplot{N3}
  \end{tabular}

  \vspace{1.5mm}
  {\small\textbf{(a)} 18 irreducible families\par}
  \vspace{2mm}

  \begin{tabular}{@{}*{6}{c}@{}}
    \onemassreducibleplot{J1} & \onemassreducibleplot{J2} & \onemassreducibleplot{J3} &
    \onemassreducibleplot{K1} & \onemassreducibleplot{K2} & \onemassreducibleplot{L1} \\[1.2mm]
    \onemassreducibleplot{L2} & \onemassreducibleplot{L3} & \onemassreducibleplot{M1} &
    \onemassreducibleplot{M2} & \onemassreducibleplot{M3} & \onemassreducibleplot{N1} \\[1.2mm]
    \onemassreducibleplot{N2} & \onemassreducibleplot{O1} & \onemassreducibleplot{O2} &
    \onemassreducibleplot{P1} & \onemassreducibleplot{P2} & \onemassreducibleplot{P3} \\[1.2mm]
    \onemassreducibleplot{Q1} & \onemassreducibleplot{Q2} & \onemassreducibleplot{R1} &
    \onemassreducibleplot{R2} & \onemassreducibleplot{R3} & \onemassreducibleplot{R4} \\[1.2mm]
    {} & \onemassreducibleplot{S1} & \onemassreducibleplot{S2} &
    \onemassreducibleplot{S3} & \onemassreducibleplot{S4} & {}
  \end{tabular}

  \vspace{1.5mm}
  {\small\textbf{(b)} 28 reducible families\par}
  \caption{The 46 integral families used in this paper. The massive external leg is shown in bold, while the labels $1$, $2$, $3$ correspond to momenta $p_1$, $p_2$, $p_3$.}
  \label{fig:one-mass-integral-families}
\end{figure}

\newcommand{\firsthere}{\multicolumn{2}{l}{\textit{first published here}}}
\newcommand{\firstherebf}{\multicolumn{2}{l}{\textbf{\textit{first published here}}}}
\newcommand{\anc}{\checkmark}
\newcommand{\ancbf}{\pmb{\checkmark}}
\newcommand{\eqv}{$\simeq$\,}
% struts to give air around the \midrule (booktabs padding is switched off below)
\newcommand{\Tstrut}{\rule{0pt}{2.5ex}}
\newcommand{\Bstrut}{\rule[-1.1ex]{0pt}{0pt}}

\begin{table}[!tbp]
 \centering
 \footnotesize
 \setlength{\tabcolsep}{3.5pt}
 % booktabs pads its rules with vertical space, which would cut the vertical
 % rule; switch the padding off (local to this table) and use struts instead.
 \setlength{\aboverulesep}{0pt}
 \setlength{\belowrulesep}{0pt}
 %
 % ----------------------- (a) irreducible -----------------------
 \begin{subtable}[t]{0.56\textwidth}
  \centering
  \begin{tabular}[t]{@{}l cc|ll c@{}}
   & & & \multicolumn{3}{c}{published as} \\
   $\mathcal F$ & $N_{\mathcal F}$ & $N_{\mathcal F}^{\mathrm{top}}$
          & identical; \qquad & other form; & this work \Bstrut \\
   \midrule
   $A$     &  83 &  3 & \cite{Henn:2023vbd}, \cite{Gehrmann:2024tds}
                      & \eqv\cite{DiVita:2014pza}, \cite{Canko:2021xmn} & \anc \Tstrut \\[3pt]
   $B_1$   & 150 &  4 & \cite{Henn:2023vbd}, \cite{Gehrmann:2024tds} & & \anc \\
   $B_2$   & 114 &  4 & \cite{Henn:2023vbd} & & \anc \\[3pt]
   $C$     & 173 &  4 & \firsthere & \anc \\[3pt]
   $D$     & 121 &  3 & \firsthere & \anc \\[3pt]
   $E_1$   & 166 &  4 & \cite{Henn:2023vbd}, \cite{Gehrmann:2024tds}
                      & \eqv\cite{Canko:2021xmn} & \anc \\
   $E_2$   & 117 &  3 & \cite{Henn:2023vbd}, \cite{Gehrmann:2024tds}
                      & \eqv\cite{Canko:2021xmn} & \anc \\[3pt]
   $F_1$   & 214 &  5 & \cite{Gehrmann:2024tds} & & \anc \\
   $F_2$   & 136 &  4 & \firsthere & \anc \\[3pt]
   $G_1$   & 254 &  6 & \cite{Gehrmann:2024tds} & & \anc \\
   $G_2$   & 174 &  4 & \firsthere & \anc \\
   $G_3$   & 176 &  4 & \firsthere & \anc \\
   $G_4$   & 212 &  5 & \firsthere & \anc \\[3pt]
   $\boldsymbol{H}$ & \textbf{371} & \textbf{19} & \firstherebf & \ancbf \\[3pt]
   $I_1$   & 277 &  8 & \firsthere & \anc \\
   $I_2$   & 305 &  8 & \cite{Gehrmann:2024tds} & & \anc \\
   $I_3$   & 333 & 12 & \firsthere & \anc \\[3pt]
   $N_3$   &  90 &  1 & & as $B_3$ \cite{Gehrmann:2024tds} & \anc \\
  \end{tabular}
  \caption{Irreducible families}
  \label{tab:family-master-counts-irred}
 \end{subtable}
 \hfill
 %
 % ------------------------ (b) reducible ------------------------
 \begin{subtable}[t]{0.40\textwidth}
  \centering
  \begin{tabular}[t]{@{}l cc|c c@{}}
   & & & \multicolumn{2}{c}{all within \cite{Canko:2021xmn, Syrrakos:2023mor} and } \\
   $\mathcal F$ & $N_{\mathcal F}$ & $N_{\mathcal F}^{\mathrm{top}}$
          & \quad in \cite{Gehrmann:2024tds}; & this work \Bstrut \\
   \midrule
   $J_1$ & 114 & 0 & & \Tstrut \\
   $J_2$ & 112 & 0 & & \\
   $J_3$ & 150 & 0 & as $B_6$ & \\[3pt]
   $K_1$ &  52 & 0 & & \\
   $K_2$ &  21 & 0 & & \\[3pt]
   $L_1$ &  89 & 0 & as $B_7$ & \\
   $L_2$ &  58 & 0 & & \\
   $L_3$ & 131 & 0 & & \anc \\[3pt]
   $M_1$ &  22 & 0 & & \\
   $M_2$ &  19 & 0 & & \\
   $M_3$ &  80 & 0 & & \anc \\[3pt]
   $N_1$ &  36 & 0 & & \\
   $N_2$ &  32 & 0 & & \\[3pt]
   $O_1$ &  23 & 0 & & \\
   $O_2$ &  50 & 0 & & \\[3pt]
   $P_1$ &  52 & 0 & & \\
   $P_2$ &  70 & 0 & as $B_5$ & \\
   $P_3$ & 143 & 0 & as $B_4$ & \anc \\[3pt]
   $Q_1$ &  86 & 0 & & \\
   $Q_2$ &  83 & 0 & & \\[3pt]
   $R_1$ &  48 & 0 & & \\
   $R_2$ & 100 & 0 & & \anc \\
   $R_3$ &  47 & 0 & & \\
   $R_4$ &  45 & 0 & & \\[3pt]
   $S_1$ &  36 & 0 & & \\
   $S_2$ &  32 & 0 & & \\
   $S_3$ &  32 & 0 & & \\
   $S_4$ &  79 & 0 & & \\
  \end{tabular}
  \caption{Reducible families}
  \label{tab:family-master-counts-red}
 \end{subtable}
 \caption{The number of master integrals per family including subsectors, $N_{\mathcal F}$, and the size of the subset appearing in the top sector 1023, $N_{\mathcal F}^{\mathrm{top}}$, are given in the first columns. 
 Additionally, we indicate which integrals have been published in the literature either with the same propagator definition, under a different name, or in a different but mapping-equivalent form.
 The reducible families are all effectively contained in the single planar family $PL$ of
 \cite{Canko:2021xmn, Syrrakos:2023mor}; some additionally appear
 individually in \cite{Gehrmann:2024tds} under different names.
 The last column indicates our choice of families published in the ancillary files to this work. Together they span the full set of three-loop four-point integrals with one off-shell leg.
 }
 \label{tab:family-master-counts}
\end{table}

\subsection{Master integrals}
The integrals in \cref{eq:scalar-integral-definition} obey
integration-by-parts (IBP) identities~\cite{Chetyrkin:1981qh,Laporta:2000dsw},
which follow from the vanishing of total derivatives in dimensional
regularisation,
\begin{equation}
  0=\int\prod_{r=1}^{3}\diff^d k_r\,
  \frac{\partial}{\partial k_\ell^\mu}
  \left[v^\mu\prod_{j=1}^{15}
  \bigl(Q_j^{\mathcal F}\bigr)^{-a_j}\right]\,,
  \qquad
  v\in\{k_1,k_2,k_3,p_1,p_2,p_3\}\,.
  \label{eq:ibp-identities}
\end{equation}
Discrete shifts of loop momenta can give rise to additional relations, between integrals of the same sector or between integrals of different sectors.
Finally, in some cases, we identify further independent relations between integrals of different sectors that follow from permutation symmetries in the Feynman parametric representation, i.e.\ from analysis of Symanzik polynomials.
Together, these relations reduce every
integral in a family to a finite set of master integrals,
\begin{equation}
  I^{\mathcal F}_{\boldsymbol a}
  =\sum_{i=1}^{N_{\mathcal F}}
  c_{\boldsymbol a, i}(s,t,m^2;\epsilon)\,M_i^{\mathcal F}\,,
  \qquad
  c_{\boldsymbol a, i}\in\mathbb Q(s,t,m^2,\epsilon)\,.
  \label{eq:ibp-reduction}
\end{equation}
In this work, we perform integral reductions with the codes \texttt{Kira\;3}~\cite{Lange:2025fba}, \texttt{Reduze\;2}~\cite{vonManteuffel:2012np} and \texttt{Finred} based on finite field arithmetic \cite{vonManteuffel:2014ixa, Peraro:2016wsq}.

\Cref{tab:family-master-counts} gives the number $N_{\mathcal F}$ of master integrals in each family, including all subsectors, and the number $N_{\mathcal F}^{\mathrm{top}}$ in its top-level ten-propagator sector.
It also records the correspondence with family names used in earlier publications.
Masters shared by different families contribute to each family's count, so the total number of independent masters is not equal to the sum of the various $N_{\mathcal F}$.

Among the 46 top-level families, master integrals
occur in the ten-propagator sectors of the following 18 families, dubbed \emph{irreducible} families:
\begin{equation}
 \mathfrak F_{\mathrm{irred}}=
 \{A,B_1,B_2,C,D,E_1,E_2,F_1,F_2,G_1,G_2,G_3,G_4,
 H,I_1,I_2,I_3,N_3\}\,.
 \label{eq:irreducible-family-labels}
\end{equation}
The ten-propagator sectors of the remaining 28 families are fully reducible via IBPs to subsectors, i.e.\ they contain no master integrals.
After graph isomorphisms and all six crossings (permutations of $p_1,p_2,p_3$)
are taken into account, representatives of the lower-sector masters can be
chosen from the irreducible families together with only four of the reducible
ones.
We choose the four reducible families
\begin{equation}
  \mathfrak F_{\mathrm{red}}^{\mathrm{MI}}
  =\{L_3,M_3,P_3,R_2\}\,.
  \label{eq:reducible-master-family-labels}
\end{equation}
Thus the 18 irreducible families and these four reducible families, together with all their crossings, cover the complete set of master integrals.

The assignment of a subsector to a particular family is a matter of organisation rather than a physical distinction. We provide a complete partition of the sectors containing master integrals across the 46 families into equivalence classes in \texttt{sector\_partition.m} in the ancillary files.

\begin{figure*}[t]
  \centering
  \includegraphics[width=\textwidth]{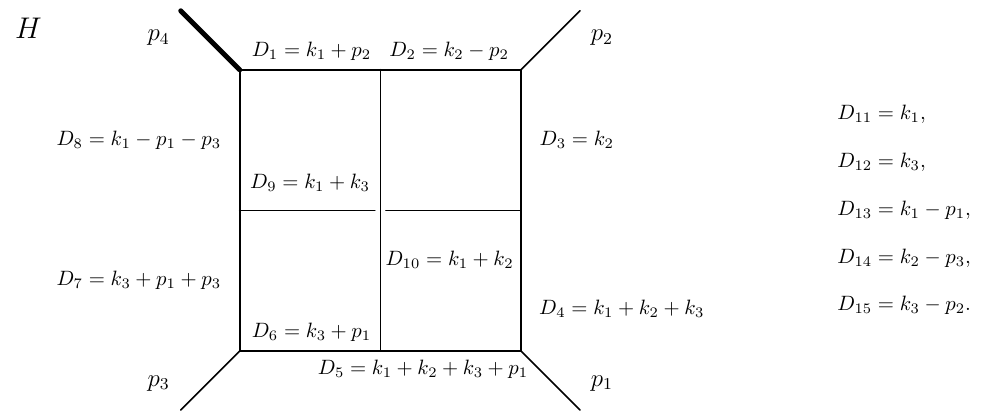}
  \caption{The propagator assignment for integral family $H$.
  The bold external line carries the off-shell momentum \(p_4\).
  The auxiliary inverse propagators \(D_{11},\ldots,D_{15}\)
  are listed on the right.}
  \label{fig:family-H-propagators}
\end{figure*}

Family $H$ has the  largest number of master integrals (371) of all families, and its
19-dimensional top sector is also the largest sector among all families
considered here.
Every other top sector contains at most 12 masters, and every proper subsector at most ten. 
Its unusually large top-sector block makes $H$ the most demanding family in the calculation, see also~\cite{Guan:2025awp,Guan:2026igf}. \Cref{fig:family-H-propagators} gives its momentum routing and propagator assignment.

\subsection{Differential equations}
For each family, we collect the master integrals into a vector $\boldsymbol M$, suppressing the family label and factoring out their overall scale dependence.
Differentiating with respect to the independent ratios $x$ and $y$ at fixed $m^2$, with $z=1-x-y$, and reducing the derivatives by
the same IBP identities gives the differential-equation system~\cite{Kotikov:1991pm,
Remiddi:1997ny,Gehrmann:1999as}
\begin{equation}
  \diff\boldsymbol M(x,y;\epsilon)
  =A(x,y;\epsilon)\,\boldsymbol M(x,y;\epsilon)\,.
  \label{eq:master-differential-equation}
\end{equation}
Here $A$ is a matrix-valued one-form in $x$ and $y$, whose entries depend explicitly on the regulator $\epsilon$.
To solve this system, we seek a rational or algebraic change of master-integral basis
$\boldsymbol J=T\boldsymbol M$ for which the system assumes the \emph{canonical
form}~\cite{Henn:2013pwa}
\begin{equation}
  \diff\boldsymbol J(x,y;\epsilon)
  =\epsilon\,\diff\widetilde A(x,y)\,\boldsymbol J(x,y;\epsilon)
  =\epsilon\sum_{\alpha}A_\alpha\,
  \diff\log W_\alpha(x,y)\,\boldsymbol J(x,y;\epsilon)\,,
  \label{eq:canonical-differential-equation}
\end{equation}
where the matrices $A_\alpha$ have constant rational entries and the functions
$W_\alpha$ form the \emph{alphabet} of the function space.
The solution is obtained by applying the path-ordered exponential of the canonical connection to a boundary vector.
Its expansion in $\epsilon$ is expressed in terms of iterated integrals, as discussed in \cref{sec:iterint-solution}.
The construction of a canonical basis for the present three-loop families is the subject of the next section.

\section{Constructing integrands with constant leading singularities}
\label{sec:dlog-strategy}
To construct a canonical basis, we follow the strategy
of~\cite{Arkani-Hamed:2010pyv,Henn:2013pwa,Henn:2020lye} and select master
integrals by analysing their four-dimensional integrands. For each integral
family, we look for numerator insertions such that the integrand has only
logarithmic poles and all of its leading singularities, i.e.\ its fully
localised residues, are independent of the external kinematics. 
It is conjectured, though not proven, that these conditions are sufficient
for a canonical basis. We therefore always verify a posteriori that
the resulting integrals satisfy differential equations in canonical form for
generic~$\epsilon$.

A systematic way to carry out this analysis is the iterated-residue method
introduced in~\cite{Henn:2020lye}, where it was also first automated in the
\texttt{DlogBasis} package. The method generates an ansatz for the numerator,
imposes the linear constraints that eliminate higher-order poles, computes the
leading singularities of the surviving integrands, and from these constructs
candidates with unit leading singularities. We adopt this general strategy and address two complications that appear in some of the present families -- intermediate residues become algebraic, and the linear systems
grow large. For some candidates in the $H$ family, see \cref{fig:family-H-propagators}, we
first perform the analysis on a suitably chosen generalised cut and then lift
the result to the full, uncut family.

The choice of parametrisation is a key ingredient for the analysis of candidate loop integrands.
It fixes the number of integration variables, determines how directly the
original propagators are expressed in terms of the new variables (and thus
how easily they can be localised), and it sets the functional form of the
integrand, which governs the technical complexity of computing the iterated
residues. 
A common choice is the Baikov
representation~\cite{Baikov:1996iu,Frellesvig:2017aai} at fixed even
dimension, $d=2,4,6$, etc.
For the
families considered here, however, this representation involves
fifteen variables and, for every even~$d$, contains the Baikov polynomial
raised to a half-integer power, which complicates the form of the integrand.
For several families, most notably
family~$H$, we could not identify a complete set of candidates in this
way.\footnote{One possible reason is that different parts of the relevant graphs
appear to call for an analysis at different values of~$d$. Family~$H$, for
instance, can be viewed as a two-loop six-point double-pentagon integral
integrated over a one-loop box, where the former is naturally analysed in
$d=6$ and the latter in $d=4$.}
For some of the harder families, such as~$I_3$, we did succeed in
constructing canonical candidates using an improved version of the Baikov
representation, described in appendix~\ref{app:improved-baikov-I3}. 

The strategy that ultimately
succeeded for the hardest cases is instead based on
a four-dimensional parametrisation of the loop momenta,
following~\cite{Henn:2020lye}. In particular, we employ the spinors of the three massless external momenta $p_1,p_2,p_3$ to
choose the transverse null vectors
\begin{equation}
 e_+^\mu=\frac{1}{2}[2|\gamma^\mu|1\rangle
              \frac{\langle23\rangle}{\langle13\rangle}\,,
 \qquad
 e_-^\mu=\frac{1}{2}[1|\gamma^\mu|2\rangle
              \frac{\langle13\rangle}{\langle23\rangle}\,.
 \label{eq:transverse-spinor-vectors}
\end{equation}
Here $p_i^\mu=\tfrac12[i|\gamma^\mu|i\rangle$. The ratios of angle
brackets make $e_+$ and $e_-$ invariant under little-group transformations of
the external spinors, so scalar products can be expressed in terms of
$s,t,m^2$. Their normalisation gives
\begin{equation}
 p_1\cdot p_2=\frac{s}{2}\,,\qquad
 e_+\cdot e_-=-\frac{s}{2}\,,\qquad
 p_i\cdot e_\pm=0\quad(i=1,2)\,,\qquad e_\pm^2=0\,.
 \label{eq:transverse-null-basis}
\end{equation}
With $u=m^2-s-t$, the remaining external momentum obeys
$2p_3\cdot e_+=u$ and $2p_3\cdot e_-=t$, or equivalently
\begin{equation}
 p_3=\frac{u}{s}p_1+\frac{t}{s}p_2
       -\frac{t}{s}e_+-\frac{u}{s}e_-\,.
 \label{eq:p3-null-basis}
\end{equation}
The three loop momenta are then parametrised as
\begin{equation}
\begin{aligned}
 k_1&=a_1p_1+a_2p_2+a_3e_++a_4e_-\,,\\
 k_2&=b_1p_1+b_2p_2+b_3e_++b_4e_-\,,\\
 k_3&=c_1p_1+c_2p_2+c_3e_++c_4e_-\,.
\end{aligned}
 \label{eq:abc-loop-parametrisation}
\end{equation}
With this parametrisation, all inverse propagators are
polynomials in the twelve dimensionless coordinates
$\boldsymbol z=(a_1,\ldots,a_4,b_1,\ldots,b_4,c_1,\ldots,c_4)$, for example $k_1^2=s(a_1a_2-a_3a_4)$.

In a given sector $S$, we search for canonical candidates starting from the numerator ansatz
\begin{equation}
 \Omega(\boldsymbol\alpha ; \boldsymbol z )
 =\mathcal J\,
 \frac{\sum_{i=1}^{n}\alpha_i N_i(\boldsymbol z)}
 {\prod_{j\in S}D_j(\boldsymbol z)}\,
 \diff z_1\wedge\cdots\wedge\diff z_{12}\,,
 \label{eq:dlog-ansatz}
\end{equation}
where $N_i$ are monomials in the $\boldsymbol z$ subject to a bound on their degree
determined by the behaviour of the corresponding integrals at infinity~\cite{Henn:2020lye}.
The Jacobian $\mathcal J = s^6$ is independent of the loop coordinates. 
Monomials containing a denominator
$D_j$, $j\in S$, cancel that propagator and therefore include contributions
from lower sectors. The ansatz is rational in the loop coordinates and
the coefficients $\alpha_i$ depend only on external invariants and may include
algebraic normalisations; they are determined by the logarithmic and residue conditions below.

We analyse the integrands iteratively, one variable at a time. 
In each variable, a logarithmic singularity is locally a simple pole. At each step of our iteration, assuming that the integrand has a single
pole at $z=0$, its singular part close to the singularity
takes the form
\begin{align}
\frac {\diff z}{z}\wedge\omega\,,
\end{align}
where $\omega$ is regular in $z$ at $z=0$ but may 
still have singularities in the remaining integration variables.
If all successive iterations in all variables only expose single poles, 
including at infinity, we say that the integrand has only logarithmic singularities and is
in d-log form.
In our case, there are twelve integration variables, and so to fully analyse the integrand
we must consider
twelve layers of iterated residues in all variables $\boldsymbol z$. 
The leading singularities of the integrand are then all the fully localised contour integrals
\begin{equation}
 \operatorname{LS}_{\Gamma}(\Omega)
 =\frac{1}{(2\pi i)^{12}}\oint_{\Gamma}\Omega,
 \label{eq:leading-singularity-definition}
\end{equation}
where $\Gamma$ runs over all iterated contours around the single poles exposed by this iterated analysis.
An integrand with constant leading singularities has numerical
residues when evaluated along these contours. 
In particular, we say that an integrand has \emph{unit} leading singularities when every such non-zero residue is $\pm 1$~\cite{Cachazo:2008vp,Arkani-Hamed:2010pyv}.

In summary, given an ansatz as in~\cref{eq:dlog-ansatz}, 
finding integrands with unit leading singularities requires three main steps: 
\begin{enumerate}
    \item Fixing the coefficients of the ansatz to make sure that there are no higher poles.
    \item Collecting all leading singularities associated to the remaining single poles. This is the heaviest step and its feasibility strongly depends on the parametrisation and on the ordering chosen to compute the residues iteratively.
    \item Solving a linear system of equations to determine how many integrals with unit leading singularities exist.
\end{enumerate}

We describe here the main steps of the algorithm, highlighting the improvements we implemented.
\paragraph{Higher poles and variable selection.}
Higher-order poles are excluded by linear constraints on the numerator
coefficients as the residue calculation proceeds. Without loss of generality, let us consider a pole at the origin, $z=0$.\footnote{Infinity can be treated in the same way after reparametrising $w=1/z$.}
Near such a pole, the integrand can be written as
\begin{equation}
 \omega(\boldsymbol\alpha)
 =\sum_{k=1}^{p}\frac{\diff z}{z^k}
       \wedge\omega_k(\boldsymbol\alpha)+\omega_{\rm reg}.
 \label{eq:nonlog-laurent}
\end{equation}
Logarithmic behaviour requires $\omega_k(\boldsymbol\alpha)=0$ for $k>1$
which yields linear equations for $\boldsymbol\alpha$. 
In particular, 
for a rational one-form $P(z)\diff z/Q(z)$, with common factors cancelled, it is easy to see that the condition at infinity becomes $\deg_z P\leq\deg_z Q-1$. 

At each iteration, once the higher poles are removed, the constraints are applied to the full
ansatz. The next layer of residues is then computed
by scoring all
 remaining variables according to their denominator degrees, the expected
 number and size of residue terms, and their root dependence. As a first difference 
 from \texttt{DlogBasis}, these scores are
 reassessed after imposing each new numerator constraint and after taking each new residue,
 to guarantee that at each layer always the simplest variable is chosen.
 For the determination of the scores, variables which only appear linearly in all denominator
 factors are preferred, and their simple-pole residues are
 evaluated directly without performing a full partial-fraction decomposition. Linear denominators in a variable $z$ are understood as arbitrary expressions of the form 
 $$\frac{\mathcal{N}\, \diff z}{(z-a_1)^{b_1}...(z-a_n)^{b_n}}$$
 where $\mathcal{N}$ is a polynomial, the $a_j$ are rational functions of all other variables involved, and $b_j$ are integer powers.

\paragraph{Intermediate algebraic roots.}
If it is not possible to factorise the denominator into factors that are all linear in at least one variable on the field of rational functions, a fully linear choice (as preferred above) is not possible, and one is forced to take residues of quadratic factors. In this way, one leaves the field of rational functions
and introduces square roots into the residue calculation.
For instance, if $q(z)=Az^2+Bz+C$ and $f$ is regular at its two distinct
zeros, then
\begin{equation}
 \underset{z=z_\pm}{\operatorname{Res}}
 \frac{f(z)\,\diff z}{q(z)}
 =\frac{f(z_\pm)}{\pm\sqrt{\mathcal D}}\,,\qquad
 z_\pm=\frac{-B\pm\sqrt{\mathcal D}}{2A}\,,\qquad
 \mathcal D=B^2-4AC\,.
 \label{eq:quadratic-residue}
\end{equation}
Here $A,B,C$ depend on the not-yet-eliminated variables. The new square root
therefore usually depends on loop coordinates. A further quadratic residue
can introduce a second root whose coefficients contain the first, increasing
both algebraic complexity and expression size.

To address this problem, \texttt{DlogBasis} searches for changes of variables that reduce the degree under the root or permit its rationalisation. However, for the families considered here these transformations are not sufficient, and many intermediate terms cannot be brought to d-log form, which prevents the calculation from being completed. In particular, we find that a brute-force rationalisation of a square root often introduces denominator factors of degree higher than three, whose residues are too costly to evaluate.

Interestingly, we find that it is often easy to treat these cases by 
finding variable transformations that remove the dependence on one of the variables from
the square root. Using the names $x,y,z$ for three generic integration variables, we implement this in practice by extending the heuristic search
to linear shifts, ratio substitutions, rescalings that
separate a common scale, and substitutions of the form $x\mapsto x\pm yz$.
After each transformation, candidates are assessed by denominator degrees, 
root dependence, and expression size. 
Importantly, when a root remains that depends on all variables, 
and at least on one of them at most quadratically,
we never rationalise the root, but instead always compute
its residues in the quadratic variable directly on its algebraic cover.

The transformation search is organised by increasing complexity, through a
finite library of coordinate charts.  The following templates are tested in order:
\begin{enumerate}
 \item null-direction shifts,
 \[
  x=u+\lambda y;
 \]
 \item directed ratio charts,
 \[
  x=yr;
 \]
 \item triangular product shears,
 \[
  x=u\pm yz;
 \]
 \item projective common-scale charts in which either two coordinates,
 \[
  x=p\xi,\qquad y=p\eta,
 \]
 or three coordinates,
 \[
  x=p\xi,\qquad y=p\eta,\qquad z=p\zeta,
 \]
 are rescaled by a retained pivot $p$.
\end{enumerate}
%The Jacobians of the ratio and common-scale charts are respectively $y$,
%$p^2$ and $p^3$, while the two triangular shifts have unit Jacobian.
All admissible choices of the participating coordinates are tested; for the
product shears both signs are included, and for the common-scale charts every
possible pivot and subset of rescaled coordinates is considered.
If this finite chart search still fails to produce a manageable residue
ordering, we employ a more general single-coordinate shift,
\[
 x=u+
 \frac{
  c_0+\sum_i c_i y_i
  +\sum_{i<j}c_{ij}y_i y_j
  +\sum_i d_i y_i^2
 }{s},
 \qquad
 s\in\{1,y_1,\ldots,y_m\},
\]
where the $y_i$ are the remaining integration variables.  The coefficients
are determined algebraically by requiring the transformed radicand to become
at most quadratic in at least one possible residue variable.  Although only
$x$ is replaced, the transformation can depend on several other
coordinates. 

Finally, as a special case, 
when only two variables remain and the direct algebraic treatment fails, the
program also attempts a separating-root chart.  Given two factors $f_i$ and
$f_j$ of the radicand that are linear in one variable, it introduces their
ratio $r=f_j/f_i$, solves this relation for that variable, and tests the
additional substitutions
\[
 y=r^m t,\qquad m\in\{-2,-1,1,2\}.
\]
Such a chart is retained only if it reduces the square-free radicand to a
function of a single new coordinate.

Importantly, since the search is heuristic, it is not guaranteed that the classes of changes of variables currently considered cover also other, more general problems. For this reason, in case a term remains unresolved after all classes of transformations are tried, we implemented a fail-safe fall-back which saves the term in a queue of unresolved terms and moves to the next one.
This allows us not to lose all steps already completed and to reconsider the failed terms separately at the end of the calculation. For the problem under study, the above templates are enough to resolve all residues.

As a simple example, consider the three-form
\begin{equation}
 \omega_{\rm ex}
 =\frac{2\,\diff x\wedge\diff y\wedge\diff z}
 {(x^2+y^2+2z^2)\sqrt{x^2+y^2+z^2}}\,.
 \label{eq:toy-root-form}
\end{equation}
A naive application of the iterated residue algorithm returns this form as unresolved due to the quadratic dependence on all variables. Instead, our variable transformation search 
tests each of the three
variables as a possible candidate to be extracted from the root. 
While the explicit variable selected depends on their internal ordering,
choosing $z$ and setting
\[
 x=z\xi\,,\qquad y=z\eta
\]
gives the Jacobian
\[
 \diff x\wedge\diff y\wedge\diff z
 =z^2\,\diff\xi\wedge\diff\eta\wedge\diff z\,.
\]
Applying the change of variable, the square root supplies one
additional power of $z$, and therefore
\begin{equation}
 \omega_{\rm ex}
 =\frac{2\,\diff\xi\wedge\diff\eta\wedge\diff z}
 {z(2+\xi^2+\eta^2)\sqrt{1+\xi^2+\eta^2}}\,.
 \label{eq:toy-root-chart}
\end{equation}
The scale dependence is thus manifestly logarithmic.  The other choices of
pivot also expose the common scale; the $z$-chart is particularly convenient
because, upon defining
\[
 R=\sqrt{1+\xi^2+\eta^2},
\]
the remaining rational denominator becomes simply $R^2+1$.

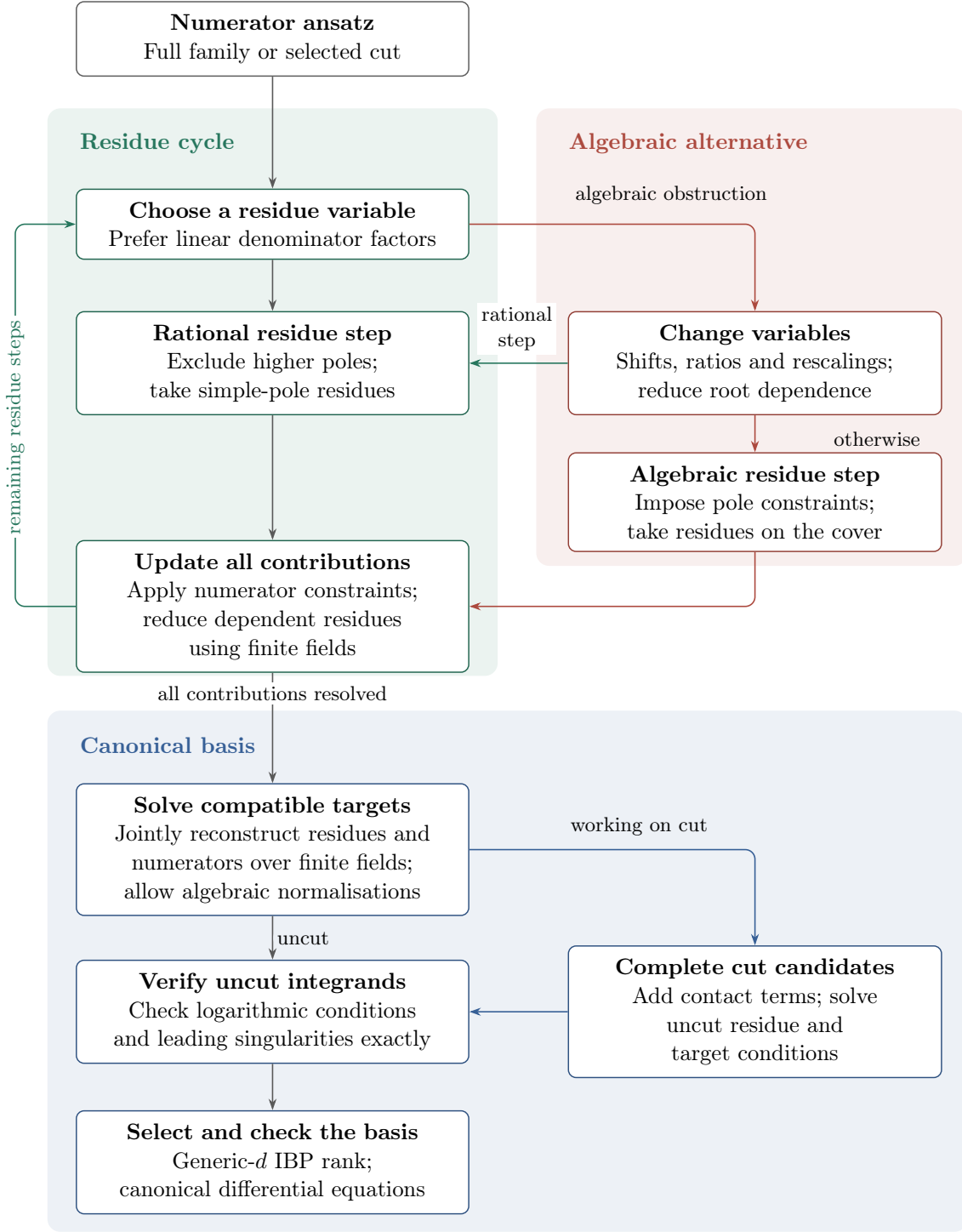
\begin{figure}[p]
\centering
\begingroup
\definecolor{residuegreen}{HTML}{26765D}
\definecolor{residuered}{HTML}{AC493E}
\definecolor{basisblue}{HTML}{365F96}
\begin{tikzpicture}[x=1mm,y=.90mm,
  methodbox/.style={draw,rounded corners=1.2mm,line width=.55pt,
    text width=57mm,align=center,inner xsep=2mm,inner ysep=2mm,
    font=\fontsize{10}{11.5}\selectfont,fill=white},
  rationalbox/.style={methodbox,draw=residuegreen!85!black},
  algebraicbox/.style={methodbox,text width=54mm,draw=residuered!85!black},
  basisbox/.style={methodbox,draw=basisblue!85!black},
  conditionalbox/.style={basisbox,text width=54mm},
  methodarrow/.style={-{Stealth[length=1.8mm,width=1.25mm]},
    line width=.6pt,draw=black!65,rounded corners=1.5mm},
  brancharrow/.style={methodarrow,draw=residuered},
  returnarrow/.style={methodarrow,draw=residuegreen},
  methodlabel/.style={font=\fontsize{9}{10.5}\selectfont,
    align=center,inner sep=.6mm,fill=white},
  areatitle/.style={anchor=west,font=\fontsize{10}{11.5}\selectfont\bfseries,
    inner sep=0pt}]

% Areas describe roles, not independent programs or fixed execution schedules.
\fill[residuegreen!9,rounded corners=2mm] (7,-12) rectangle (77,-110);
\fill[residuered!9,rounded corners=2mm] (83,-12) rectangle (150,-91);
\fill[basisblue!9,rounded corners=2mm] (7,-116) rectangle (150,-206);
\node[areatitle,text=residuegreen] at (12,-18) {Residue cycle};
\node[areatitle,text=residuered] at (88,-18) {Algebraic alternative};
\node[areatitle,text=basisblue] at (12,-122) {Canonical basis};

\node[methodbox,draw=black!65] (input) at (42,0)
  {\textbf{Numerator ansatz}\\ Full family or selected cut};
\node[rationalbox] (select) at (42,-32)
  {\textbf{Choose a residue variable}\\
   Prefer linear denominator factors};
\node[rationalbox] (rational) at (42,-56)
  {\textbf{Rational residue step}\\
   Exclude higher poles;\\ take simple-pole residues};
\node[algebraicbox] (transform) at (117,-56)
  {\textbf{Change variables}\\
   Shifts, ratios and rescalings;\\ reduce root dependence};
\node[algebraicbox] (algebraic) at (117,-80)
  {\textbf{Algebraic residue step}\\
   Impose pole constraints;\\ take residues on the cover};
\node[rationalbox] (update) at (42,-98)
  {\textbf{Update all contributions}\\
   Apply numerator constraints;\\
   reduce dependent residues\\ using finite fields};

\draw[methodarrow] (input) -- (select);
\draw[methodarrow] (select) -- (rational);
\draw[brancharrow] (select.east) -- (117,-32) -- (transform.north);
\node[methodlabel,fill=residuered!9] at (104,-27)
  {algebraic obstruction};
\draw[returnarrow] (transform.west) -- (rational.east);
\node[methodlabel] at (80,-50) {rational\\step};
\draw[brancharrow] (transform) -- (algebraic);
\node[methodlabel,fill=residuered!9] at (136,-69) {otherwise};
\draw[methodarrow] (rational) -- (update);
\draw[brancharrow] (algebraic.south) -- (117,-98) -- (update.east);

% One shared iteration loop: both branches feed the updated residue problem.
\draw[returnarrow] (update.west) -- (2,-98) -- (2,-32) -- (select.west);
\node[methodlabel,rotate=90,text=residuegreen] at (2,-67)
  {remaining residue steps};

% Only complete residue data enter the target solve.
\node[basisbox] (targets) at (42,-140)
  {\textbf{Solve compatible targets}\\
   Jointly reconstruct residues and\\ numerators over finite fields;\\
   allow algebraic normalisations};
\node[basisbox] (verify) at (42,-168)
  {\textbf{Verify uncut integrands}\\
   Check logarithmic conditions\\ and leading singularities exactly};
\node[basisbox] (basis) at (42,-194)
  {\textbf{Select and check the basis}\\
   Generic-$d$ IBP rank;\\ canonical differential equations};
\draw[methodarrow] (update) -- (targets);
\node[methodlabel] at (42,-113) {all contributions resolved};
\draw[methodarrow] (targets) -- node[methodlabel,right,fill=basisblue!9] {uncut} (verify);
\draw[methodarrow] (verify) -- (basis);

% Cut completion rejoins the main path before exact verification.
\node[conditionalbox] (completecut) at (117,-168)
  {\textbf{Complete cut candidates}\\
   Add contact terms; solve\\ uncut residue and target conditions};
\draw[methodarrow,draw=basisblue] (targets.east) -- (117,-140) -- (completecut.north);
\node[methodlabel,fill=basisblue!9] at (99,-136) {working on cut};
\draw[methodarrow,draw=basisblue] (completecut.west) -- (verify.east);
\end{tikzpicture}
\endgroup
\caption{Construction of canonical-basis candidates by iterated residues.
Both residue paths impose logarithmic pole conditions, including at infinity,
and feed the same update cycle. A change of variables returns to rational
residues when possible. Unresolved terms are retained for reconsideration;
the blue stage requires complete residue data and targets compatible with
the kinematic relations among residues.}
\label{fig:dlog-algorithm}
\end{figure}

\paragraph{Dependence reduction and unresolved contributions.}
The procedure above allows us to collect, for each layer, a large number of residues, expressed as algebraic functions which depend on the remaining integration variables and on the kinematics. We remove linearly dependent residue expressions at each iteration to avoid
repeated work. At variance with the original implementation, for large residue lists, we replace linear
algebra over the rationals or algebraic numbers with finite-field sampling. To this end, each distinct intermediate
root is temporarily represented by an independent symbol, which is enough to detect
relations in the resulting rational expressions.
Reconstructed relations are checked at additional sample points and over
several prime fields; failed checks lead to rational reduction or retention of
the expressions.
\Cref{fig:dlog-algorithm} summarises the residue calculation and the
subsequent determination and verification of the candidates.
% Main path: rational residues, shared updates, then basis construction.
% The algebraic branch returns to that path when a rational step is exposed.

\subsection{Determining constant residue vectors}
\label{sec:constant-ls-space}
After all logarithmic constraints are imposed, the ansatz in~\cref{eq:dlog-ansatz}
depends on $q$ free coefficients $\boldsymbol v=(v_1,\ldots,v_q)$, which, like
the $\alpha_i$, may depend on $s,t,m^2$. Its $n_{\rm LS}$ leading
singularities are linear in them,
\begin{equation}
 \operatorname{LS}_{\Gamma_i}(\Omega)=\sum_{j=1}^{q}A_{ij}\,v_j\,,
 \qquad i=1,\ldots,n_{\rm LS}\,,
 \label{eq:ls-linear-map}
\end{equation}
where the entries of $A$ are functions of the external invariants. A candidate
is a choice of $\boldsymbol v$ for which all leading singularities are
numbers, $A\boldsymbol v=\boldsymbol\tau$ with
$\boldsymbol\tau\in\mathbb Q^{n_{\rm LS}}$. We call $\boldsymbol\tau$ the
\emph{target} of the candidate, and say that a target is \emph{compatible} if
this system has a solution.

\texttt{DlogBasis} uses the $n_{\rm LS}$ unit vectors as
targets: it requires, in turn, one leading singularity to be equal to one and
all the others to vanish. All of these targets are compatible if and only if
$A$ has rank $n_{\rm LS}$, i.e.\ if no linear combination of the leading
singularities vanishes, even allowing coefficients that depend on the
kinematics. The dependence reduction described above, however, only removes
leading singularities that are combinations of the others with
\emph{numerical} coefficients, which are automatically constant once the others
are. A leading singularity related to the others by kinematic coefficients
cannot be removed, since requiring it to be constant is an independent
condition. Every such relation makes some of the unit targets incompatible, and
the original implementation then returns no candidates at all, even though
compatible targets may exist.

As an example, consider two free coefficients and three leading singularities
\begin{equation}
 \bigl(\operatorname{LS}_1,\operatorname{LS}_2,\operatorname{LS}_3\bigr)
 =\Bigl(\frac{v_1}{s},\,\frac{v_2}{t},\,\frac{v_1+v_2}{s+t}\Bigr).
 \label{eq:compatible-target-example}
\end{equation}
None of them is a numerical combination of the others, but
$s\,{\operatorname{LS}_1}+t\,{\operatorname{LS}_2}-(s+t)\,{\operatorname{LS}_3}=0$. A
compatible target must therefore satisfy $s\tau_1+t\tau_2-(s+t)\tau_3=0$ for all
$s,t$, that is $\tau_1=\tau_2=\tau_3$. No unit vector is compatible: for
instance, $\operatorname{LS}_1=1$ and $\operatorname{LS}_2=0$ require $v_1=s$
and $v_2=0$, and then $\operatorname{LS}_3=s/(s+t)$. Nevertheless,
$(v_1,v_2)=(s,t)$ gives $\operatorname{LS}_1=\operatorname{LS}_2=\operatorname{LS}_3=1$,
an integrand with unit leading singularities that the unit-vector approach
misses.

We therefore determine all compatible targets directly. They are the constant
vectors in the image of $A$, obtained by requiring $\boldsymbol\tau$ to be
orthogonal to the left null space of $A$ at generic
kinematic points in $s,t,m^2$, which yields linear equations for $\boldsymbol\tau$ with
rational coefficients. In the example above, their solutions are spanned by
$(1,1,1)$. The basis of this space is not unique. We
select, whenever possible, basis vectors with entries in $\{0,\pm1\}$, so that
the corresponding candidates have unit leading singularities rather than
merely constant ones. For each basis target we solve
$A\boldsymbol v=\boldsymbol\tau$ using finite-field
methods as implemented in \texttt{FiniteFlow}~\cite{Peraro:2019svx}, verify every candidate by exact substitution
into all logarithmic conditions and leading singularities, and finally select
the linearly independent candidates by generic-$d$ IBP reduction.

Leading singularities that contain the root $r$
of~\cref{eq:kinematic-gram-determinant} require candidates normalised by $r$.
Writing
\begin{equation}
 A=A_0+r\,A_1\,,\qquad r^2=m^2st\,(s+t-m^2)\,,
 \label{eq:gram-root}
\end{equation}
with $A_0,A_1$ rational in $s,t,m^2$, coefficients $\boldsymbol v$ or
$r\boldsymbol v$, with $\boldsymbol v$ rational, solve respectively
\begin{equation}
 A_0\boldsymbol v=\boldsymbol\tau\,,\quad A_1\boldsymbol v=0
 \qquad\text{or}\qquad
 A_0\boldsymbol v=0\,,\quad r^2A_1\boldsymbol v=\boldsymbol\tau\,.
 \label{eq:unit-targets}
\end{equation}

This extension was essential for both families $H$ and $I_3$: in both, the
leading singularities of the quadratic numerator ansatz obey relations with
kinematic coefficients, and the original approach returns no candidates. 
This construction was useful both to speed up the construction of the candidates and to find as many candidates with unit leading singularities as possible. An 
example for family $H$ is discussed in~\cref{sec:H-lower-sector-ls}.

\subsection{Using cuts to restrict the numerator ansatz}
\label{sec:dlog-cuts}

While the strategy discussed above can be in principle applied to the full numerator ansatz, the residue calculation might become too costly. In this case, a common strategy consists in
simplifying the problem on suitable combinations 
of generalised cuts. For a set of propagators
$C\subseteq S$, the cut integrand is the residue of~\cref{eq:dlog-ansatz} at
$D_j=0$ for all $j\in C$. With the simple propagator powers
of~\cref{eq:dlog-ansatz}, every numerator monomial that contains a cut
propagator cancels it and does not contribute on the cut. The cut calculation
therefore involves fewer integration variables and a smaller ansatz, and its
leading singularities are a subset of those of the uncut integrand. 
Applying
the construction of~\cref{sec:constant-ls-space} on the cut gives numerators
$N_C$ with constant leading singularities on the cut.

Consequently, a cut numerator determines the uncut numerator only up to additive terms that vanish on
the cut. Given a solution on a cut, we can therefore look for an uncut numerator with unit leading singularities starting from an ansatz of the form
\begin{equation}
 N=N_C+\sum_{j\in C}D_j\,R_j\,,
 \label{eq:cut-completion}
\end{equation}
where the \emph{contact terms} $D_jR_j$ are built from all monomials of the
ansatz that contain at least one cut propagator. 
The coefficients of the contact terms are determined by requiring that the
uncut integrand has
only logarithmic singularities and that all its other leading singularities
are constant as well. This is again a linear problem of the type discussed
in~\cref{sec:constant-ls-space}, and we solve it in the same way. In practice, 
we keep the coefficients of the monomials of $N_C$ as free parameters during
the uncut residue calculation and fix them only in this final linear system,
see~\cref{sec:hfamily}.

\section{Application to the \texorpdfstring{$H$}{H} family}
\label{sec:hfamily}
In this section we detail the construction of a canonical basis for the top
sector of family~$H$, the most complicated family considered in this work.
A common strategy is to proceed top-down: one first constructs candidates on
the maximal cut, where the residue computation is simplest, and then lifts
them to uncut integrands by releasing the cut propagators and adding contact
terms, i.e.\ numerator terms proportional to the released propagators. Since
contact terms vanish on the cut, they cannot alter the cut integrand. They
can only cancel higher-order poles and non-constant leading singularities
that appear away from the cut, and within the degree bound of the ansatz they
may be unable to do so. In that case, the on-cut numerator must be changed as
well, and only particular linear combinations of good cut candidates admit a
completion. 
In other words, a set of candidates that is complete on a given cut, especially 
if this is the maximal cut, need
not be completable element by element. This is precisely what happens for
family~$H$, and it is part of the reason why constructing a canonical basis in this case is very non-trivial.

In particular, starting from a cubic ansatz on the maximal cut
of the $H$ top sector readily yields $40$
representatives with unit leading singularities whose generic-$d$ IBP image
has rank~$19$, matching the required number of master integrals. Several of
these, however, cannot be completed individually, and completing the others
requires increasingly complicated linear combinations of cut candidates,
with correspondingly complicated coefficients for the contact terms.
We therefore take a different route and determine the on-cut numerators and
the contact terms simultaneously.

\subsection{Quadratic numerators in the uncut parametrisation}
As a first step, we consider a quadratic ansatz, taking the common denominator $D_1\cdots D_{10}$ and allowing all
monomials of degree at most two in the fifteen inverse propagators. This
gives $136$ numerator parameters, including the corner integral and terms
that cancel propagators, i.e.\ candidates for the subtopologies. By employing the method described in \cref{sec:dlog-strategy} to solve the logarithmic constraints and the
constant-target equations, we find $109$ explicit 
integrands with constant leading singularities.
Of these candidates, $27$ contain contributions with all ten propagators
present. Their top-sector parts involve $20$ distinct integrals with
non-trivial numerators and no raised propagator powers, whose generic-$d$ IBP rank is nevertheless only $15$, leaving four canonical directions to still be determined. 

\subsection{Cubic numerators from a double cut}
To determine the remaining four master integrals, we choose an intermediate cut,
balancing two competing effects. Cutting more propagators simplifies the
residues but leaves more of the numerator to be fixed by contact terms,
making obstructions of the kind described above more likely. On the double
cut $D_3=D_6=0$, the residue computation remains feasible and the resulting
cubic candidates can be lifted to uncut integrands.
Indeed, this double cut admits a particularly
simple rational parametrisation while leaving both central crossed
propagators uncut, see~\cref{fig:family-H-propagators}. In the coordinates of
\cref{eq:abc-loop-parametrisation}, the cut equations are solved by
\begin{equation}
 b_1=\frac{b_3b_4}{b_2}\,,\qquad
 c_2=\frac{c_3c_4}{1+c_1}\,.
\end{equation}
The two equations are independent and linear in the eliminated variables,
so imposing the cut introduces no additional square roots. Moreover, the propagators
$D_9=(k_1+k_3)^2$ and $D_{10}=(k_1+k_2)^2$ remain present,
retaining the couplings between all three loop momenta.
By employing the algorithm described in \cref{sec:dlog-strategy} we find that the  cubic residue calculation on this cut supplies constant-leading-singularity
candidates whose generic-$d$ top-sector IBP image has rank $19$, thus spanning the full space.
To find the missing 4 canonical candidates and complete the 15 integrals determined on the quadratic ansatz, we start by identifying 4 of the cubic cut candidates which are linearly independent of the 15 quadratic ones under IBPs. 
For each chosen cut numerator $N_{\rm seed}$, we
retain its distinct monomials $S_i$ with 
their fixed coefficients and add
monomials $C_k$ that vanish when $D_3=D_6=0$:
\begin{align}
 &N(\boldsymbol\beta,\boldsymbol\gamma)
 =\sum_i\beta_i S_i+\sum_k\gamma_k C_k,
 & C_k&\in\langle D_3,D_6\rangle,\qquad \deg C_k\leq3,
 \notag\\
 &N(\boldsymbol\beta,\boldsymbol\gamma)\big|_{D_3=D_6=0}
 =N_{\rm seed}\big|_{D_3=D_6=0}.
 \label{eq:ideal-completion}
\end{align}
We  then derive the uncut logarithmic conditions and compute all leading singularities keeping the $\gamma_k$ as generic coefficients. For each candidate, we find that it is possible to find a solution for the $\gamma_k$ which removes all non-logarithmic poles and which allows us to normalise all leading singularities to 
$\{0,1,-1\}$. This allows us to complete the basis and find $19$
IBP-independent candidates with unit leading singularities on all cuts.

\subsection{Constant residue integrands in lower sectors}
\label{sec:H-lower-sector-ls}

The procedure described above fixes the $19$ candidates with a non-trivial
projection onto the top sector. All \emph{sub}sectors of family~$H$ could in
principle be covered by canonical integrals already constructed for the other, simpler
families, mapped onto $H$ through sector relations and kinematic crossings.
However, the form of
the differential equation (which has impact on the complexity of the solutions, boundary conditions and numerical evaluation) depends on how the integrals across the entire
sector tree are chosen relative to each other. Indeed, assembling the
subsector basis from other families in this way produces canonical
differential equations with large rational coefficients, which suggests that
the resulting basis is unnatural for family~$H$.

We therefore generate novel subsector candidates within family~$H$ itself, by
repeating the uncut construction with a full quadratic ansatz for each of its
ten nine-propagator subsectors. This yields candidates for all sectors down to
seven propagators. For instance, omitting $D_3$, the logarithmic constraints
leave $74$ free coefficients and $76$ leading singularities, of which only
$74$ are independent. The two relations among them have kinematic
coefficients, so the compatible targets must be determined as in
\cref{sec:constant-ls-space}. This gives $70$ rationally normalised
candidates and one normalised by~$r$. Repeating the same steps for all ten subsectors, 
we obtain
$741$ candidates, with substantial overlap. Only four of them are root-normalised,
and $r$ is the only root that appears in any normalisation.
At the level of linearly independent candidates, only two integrals from different eight-propagator topologies require a normalisation with the root $r$.

In order to find a particularly compact form for the differential equations, we proceed as follows.
We supplement this set with candidates from subsectors of other families and their kinematic crossings.
Having assembled all possible candidates for canonical integrals for the top sector and all of its subsectors, we sort them according to the number of integrals their definition contains.
With this ordering, we perform IBP reduction to determine a linearly independent set of candidates with preferably few terms in their definition.
This procedure substantially reduces the size of the differential equation and the involved rational numbers.

\section{Results}
\label{sec:results}
\subsection{Canonical bases and alphabet}
\label{sec:canonical-bases}

Applying the construction discussed in the previous two sections,
we obtain canonical bases for all 22 families in
$\mathfrak{F}_{\mathrm{irred}}\cup\mathfrak{F}_{\mathrm{red}}^{\mathrm{MI}}$.
For each family $\mathcal{F}$  independently, the vector of canonical integrals
$\boldsymbol{J}_{\mathcal{F}}$ satisfies
\begin{equation}
  \mathrm{d}\boldsymbol{J}_{\mathcal{F}}
  = \epsilon\sum_{\alpha=1}^{20}A_{\mathcal{F},\alpha}
    \,\mathrm{d}\log W_\alpha\,\boldsymbol{J}_{\mathcal{F}}\,,
  \label{eq:cde}
\end{equation}
where the matrices $A_{\mathcal{F},\alpha}$ have rational numerical entries.
This includes the complete 371-dimensional system of family~$H$ and,
in particular, its 19-dimensional top-sector block.
The canonical integrals together with their differential equations are
provided in the electronic ancillary files.

All systems close on the same 20-letter alphabet. To present it
compactly, we use $z = 1-x-y$ and introduce the dimensionless version of the square root
\begin{equation}
  \rho = \frac{r}{m^4} = \sqrt{-xyz}\,.
  \label{eq:rho}
\end{equation}
The six linear letters are
\begin{equation}
  \mathbb{A}_{\mathrm{lin}}
  = \{l_1, l_2, l_3, l_4, l_5, l_6\}
  = \{x,\, y,\, z,\, 1-x,\, 1-y,\, 1-z\},
  \label{eq:linletters}
\end{equation}
and two independent algebraic letters may be chosen as
\begin{equation}
  \mathbb{A}_{\mathrm{alg}}
  = \{l_7, l_8\}
  = \left\{\frac{xz - \rho}{xz + \rho},\;
           \frac{xy - \rho}{xy + \rho}\right\}.
  \label{eq:algletters}
\end{equation}
The third permutation, $(yz-\rho)/(yz+\rho)$, equals the inverse of the
product $l_7 l_8$ and therefore contributes no independent logarithmic
differential. The remaining twelve letters are rational and at most
quadratic in $x$ and $y$:
\begin{equation}
  \begin{aligned}
    \mathbb{A}_{\mathrm{quad}}
    = \{&l_9, l_{10}, l_{11}, l_{12}, l_{13}, l_{14},
          l_{15}, l_{16}, l_{17}, l_{18}, l_{19}, l_{20}\} \\
    = \{&1-(1+x)y,\; 1-x(1+y),\;
          x - y + y^2,\; y - x + x^2, \\
       &(1-x)^2 - y,\; (x+y)^2 - y,\;
          (x+y)^2 - x,\; (1-y)^2 - x, \\
       &x + xy + y^2,\; y + xy + x^2,\;
          (1-y)^2 + xy,\; (1-x)^2 + xy\}.
  \end{aligned}
  \label{eq:quadletters}
\end{equation}
In~\cite{Gehrmann:2024tds}, letters $l_{11}-l_{16}$ were dubbed \emph{parabolic}, while $l_{9},l_{10},l_{17}-l_{20}$ were called \emph{hyperbolic} due to the curves they describe in the $(x,y)$ plane.

In conclusion, $\mathbb{A} = \mathbb{A}_{\mathrm{lin}} \cup \mathbb{A}_{\mathrm{alg}}
\cup \mathbb{A}_{\mathrm{quad}}$ with $|\mathbb{A}| = 20$ total letters closed under all permutations of the massless external legs.
Up to permutations, overall constant factors, and inversions of
individual letters, this coincides with the alphabet identified
in~\cite{Gehrmann:2024tds} and quoted in~\cite{Guan:2026igf}.
The six linear letters constitute the complete alphabet through two loops
and for all planar topologies at three loops; the algebraic and quadratic
letters first arise in non-planar topologies at three loops.

\subsection{Boundary conditions}
\label{sec:boundary-conditions}

The physically relevant solutions to the canonical differential equations are fully determined only after we provide boundary vectors
through transcendental weight six. We fix them using the method of
ref.~\cite{Gehrmann:2024tds}, which avoids constructing a global
functional representation prior to fixing the boundary constants.
Regularity at the pseudo-thresholds $x=1$, $y=1$, and $z=1$, together
with the prescribed asymptotic behaviour as $x$, $y$, or $z$ tends to
zero, relates the unknown constants to elementary one-scale integrals.
The resulting boundary data are transported to a common base point along
one-dimensional kinematic slices on which the alphabet linearises.
At intersections of singular surfaces the matching path is regularised
by a blow-up: a change of coordinates that renders each two-dimensional limit unambiguous. This procedure fixes all
boundary constants for every family.

We denote the regularised boundary vector at the corner $x = y = 0$ by
\begin{equation}
  \boldsymbol{b}_{\mathcal{F}}(\epsilon)
  = \underset{x,y\to 0}{\operatorname{Reg}}\,
    \boldsymbol{J}_{\mathcal{F}}(x,y;\epsilon)\,.
  \label{eq:bvec}
\end{equation}
Its coefficient at order $\epsilon^w$ has uniform transcendental weight $w$.
Through the order required here, the boundary constants
belong to the spaces spanned by
\begin{equation}
  1,\quad \pi^2,\quad \zeta_3,\quad \pi^4,\quad
  \{\pi^2\zeta_3,\,\zeta_5\},\quad
  \{\pi^6,\,\zeta_3^2\}
  \label{eq:transcendental_basis}
\end{equation}
at weights zero, two, three, four, five, and six, respectively; the
weight-one constants vanish in our normalisation.
As a representative example, the first component of the $H$-family
boundary vector reads
\begin{equation}
  \begin{aligned}
    b_{H,1}(\epsilon) = {}&
      1 - \frac{\pi^2}{4}\epsilon^2
        - 29\zeta_3\,\epsilon^3
        - \frac{71\pi^4}{160}\epsilon^4 \\
      &{} + \left(\frac{29\pi^2\zeta_3}{4}
               - \frac{1263\zeta_5}{5}\right)\epsilon^5
        + \left(-\frac{11539\pi^6}{24192}
               + \frac{841\zeta_3^2}{2}\right)\epsilon^6
        + \mathcal{O}(\epsilon^7)\,.
  \end{aligned}
  \label{eq:bH1}
\end{equation}
The complete $H$ boundary vector has 371 components, many of comparable
length. We provide all boundary vectors, including those for all other
families and their kinematic crossings, in the electronic ancillary files.

\subsection{Solution in terms of Chen iterated integrals}
\label{sec:iterint-solution}

The canonical form in~\cref{eq:cde} admits a formal solution given by the
path-ordered exponential
\begin{equation}
  \boldsymbol{J}_{\mathcal{F}}(\vec{x},\epsilon)
  = \mathbb{P}\exp\!\left(\epsilon\int_{\gamma}\mathrm{d}\tilde{A}\right)
    \boldsymbol{b}_{\mathcal{F}}(\epsilon)\,,
  \label{eq:pathexp}
\end{equation}
where $\gamma$ connects the base point $\vec{x}_0 = (0,0)$ to the
kinematic point $\vec{x} = (x,y)$.
Expanding in powers of $\epsilon$, the solution is naturally expressed
through Chen iterated integrals~\cite{Chen:1977oja}, with integration
kernels drawn from the alphabet letters:
\begin{equation}
  I(\omega_1,\ldots,\omega_n;\vec{x})
  = \int_{\gamma}\omega_1\,\omega_2\cdots\omega_n\,,
  \qquad
  I(;\vec{x}) = 1\,,
  \label{eq:chen}
\end{equation}
where each $\omega_i = \omega_i(x,y)$ is taken from the alphabet defined
in \crefrange{eq:linletters}{eq:quadletters}.
The weight of the iterated integral is defined as the number of integration kernels in its index vector.
Once all products of iterated integrals with the same argument
are rewritten as sums of single iterated integrals using the shuffle product, solutions written in terms of these objects have no hidden zeros.
In other words, two expressions involving functions $I(\cdots)$ with different sets of differential forms are genuinely different, and more general expressions can be handled by considering $\mathbb{Q}$-linear relations.

We derive solutions for all master integrals of all families, including
their kinematic crossings, by expanding~\cref{eq:pathexp} in $\epsilon$
through transcendental weight~6.
From the iterated-integral representation, we can directly read off the weight
at which each letter of the alphabet first contributes.
All quadratic letters first appear at transcendental weight~4. 
For two of the newly computed families, $H$ and $D$, we find that the algebraic
letters already appear at weight~3, one lower than for the subset of families relevant for $H$+jet in the leading-colour approximation.

For four-point integrals with one off-shell leg, the following conditions on the sequence of letters appearing in the iterated integrals have been put forward in~\cite{Dixon:2020bbt, Chicherin:2020umh}, inspired by a description using the B2/C2 cluster algebras.
Referring to the indices of the letters $l_i$, they can be checked at the level of the coefficient matrices of the corresponding one-forms in the differential equations to all orders in $\epsilon$ as
\begin{align}
    A_i \cdot A_j = A_j \cdot A_i = 0\,,\qquad i,j\in\{4,5,6\}\,,i\neq j\,,
    \label{eq:old adjacency}
\end{align}
which, however, turn out not to be satisfied for individual integrals at three loops.
In addition, seven new relations have been recently proposed based on associating part of the alphabet with the cluster algebra G2~\cite{Aliaj:2024zgp}:
\begin{align}
    A_i \cdot A_j &= 0\,,\qquad (i,j)=(5,12),(5,13), (6,12), (6,13), (12,13)\,
        \label{eq:new adjacency1}
\end{align}
and
\begin{align}
    A_{12} \cdot (A_1+A_2+A_6+A_{12})&=0\,,\\
    (A_1+A_3+A_5+A_{13}) \cdot A_{13}&=0\,.
    \label{eq:new adjacency2}
\end{align}
We find that these conditions and their kinematic crossings are satisfied for all the integral families, and therefore for any resulting amplitude.

\subsection{Graded transcendental functions}
\label{sec:trident-functions}
In ref.~\cite{Gehrmann:2024tds}, the solutions of all canonical master integrals contributing to the $H$+jet amplitude at leading colour (in 12 families: $PL$ of~\cite{Canko:2021xmn} and $A, B_1,E_1,E_2,F_1,G_1,I_2,J_3,L_1,P_2$,$P_3$ in our notation) and their six kinematic crossings were computed as Laurent expansions in $\epsilon$ through functions of weight 6.
The solutions were represented as vectors of rational number coefficients with respect to the elementary transcendental \emph{tokens}: the set of iterated integrals multiplied by powers of the dimensional regulator and constants from the basis in~\cref{eq:transcendental_basis}, for example $\epsilon^5 \, \zeta_3 \, I(l_1,l_2;\vec{x})$. 
Row reduction then determined the span of the rational coefficient vectors and hence a minimal basis of $1\,282$ transcendental functions, the \emph{trident functions} \(\psi_i\), sufficient to express any amplitude built from those families.\footnote{Note that each trident function is a particular $\mathbb{Q}$-linear combination of the canonical integral solutions, and is therefore also a Laurent expansion in $\epsilon$ of uniform weight, truncated at the order containing weight-six functions. }
The construction is an adaptation of the method developed in the context of pentagon functions~\cite{Chicherin:2020oor,Chicherin:2021dyp}.

This basis was subsequently graded by separating the functions involving only the six lower-loop letters from the 93 functions containing any new three-loop letter, and by the first weight at which such a letter appears. 
At each weight, further basis rotations isolated independent groups according to the class (parabolic, hyperbolic, root) of the new letter that they contain, and according to whether their iterated-integral letter sequences satisfy a conjectured condition involving \(l_4,l_5,l_6\) put forward in~\cite{Dixon:2020bbt, Chicherin:2020umh}. 
Each step was performed as a direct-sum decomposition, so that functions assigned to one grade cannot be generated by linear combinations of functions in the others. 
Consequently, integrals and amplitudes expressed in terms of the trident functions manifest the properties that the basis was graded on with the presence or absence of certain basis elements in their decomposition.

\newcolumntype{Y}{>{\centering\arraybackslash}X}
\newcommand{\NA}{\textemdash}
\newcommand{\hasletter}{\ensuremath{\bullet}}
\newcommand{\adjy}{\ensuremath{\checkmark}}
\newcommand{\adjn}{\ensuremath{\times}}
\newcommand{\adjnew}{\ensuremath{\sim}}
\definecolor{leadingcolour}{gray}{0.93}

\begin{table}[t]
  \centering
  \footnotesize
  \setlength{\tabcolsep}{4.5pt}
  \renewcommand{\arraystretch}{1.12}
  \begin{tabularx}{\linewidth}{@{}>{\raggedright\arraybackslash}X c cccc cc@{}}
    \toprule
    Trident functions
      & Number
      & \multicolumn{4}{c}{Letter classes}
      & \shortstack{First new-\\letter weight}
      & Adjacency \\
    \cmidrule(lr){3-6}
      & & two-loop & parabolic & hyperbolic & root & & \\
    \midrule
    \multicolumn{8}{@{}l}{\itshape Previously published basis} \\
    \addlinespace[1pt]
    $\psi_{1\text{--}3}$
      & 3 & \hasletter & & & \hasletter & 4 & \adjn \\
    \rowcolor{leadingcolour}
    $\psi_{4\text{--}9}$
      & 6 & \hasletter & \hasletter & & & 4 & \adjnew \\
    $\psi_{10\text{--}18}$
      & 9 & \hasletter & \hasletter & \hasletter & \hasletter & 5 & \adjnew \\
    $\psi_{19\text{--}24}$
      & 6 & \hasletter & \hasletter & \hasletter & & 5 & \adjnew \\
    \rowcolor{leadingcolour}
    $\psi_{25\text{--}36}$
      & 12 & \hasletter & \hasletter & & & 5 & \adjnew \\
    $\psi_{37\text{--}51}$
      & 15 & \hasletter & \hasletter & \hasletter & \hasletter & 6 & \adjnew \\
    $\psi_{52\text{--}63}$
      & 12 & \hasletter & \hasletter & \hasletter & & 6 & \adjnew \\
    \rowcolor{leadingcolour}
    $\psi_{64\text{--}93}$
      & 30 & \hasletter & \hasletter & & & 6 & \adjnew \\
    $\psi_{94\text{--}111}$
      & 18 & \hasletter & & & & \NA & \adjn \\
    $\psi_{112\text{--}1282}$
      & 1171 & \hasletter & & & & \NA & \adjy \\
    \addlinespace[2pt]
    \midrule
    \multicolumn{8}{@{}l}{\itshape Additional full-colour elements} \\
    \addlinespace[1pt]
    $\psi_{1283}$
      & 1 & \hasletter & & & \hasletter & 3 & \adjnew \\
    $\psi_{1284\text{--}1285}$
      & 2 & \hasletter & & & \hasletter & 4 & \adjnew \\
    $\psi_{1286\text{--}1288}$
      & 3 & \hasletter & & & \hasletter & 5 & \adjnew \\
    $\psi_{1289\text{--}1307}$
      & 19 & \hasletter & & & \hasletter & 6 & \adjnew \\
    $\psi_{1308\text{--}1318}$
      & 11 & \hasletter & & & & \NA & \adjy \\
    \midrule
    Total & 1318 & & & & & & \\
    \bottomrule
  \end{tabularx}
  \caption{Grading of the complete set of trident functions through
  transcendental weight six. A bullet marks a letter class present in the given grade at the listed first new-letter weight. For rows containing no new letters, a $\checkmark$ means that every
  function satisfies the condition~\cref{eq:old adjacency}, whereas a $\times$ means that at least one
  function violates it. For rows containing new letters, a $\sim$ means that
  no token containing a \emph{new} letter violates this condition, whereas
  a $\times$ means that at least one such token violates it; neither symbol
  makes a statement about the part containing only $l_{1\text{--}6}$. 
  Light-grey rows are the three sets of functions that
  enter the leading-colour contributions evaluated in ref.~\cite{Gehrmann:2024tds}.}
  \label{tab:new-trident-functions}
\end{table}

In this work, we express the \emph{full} set of $23\,520$ integrals from 22 families and in 6 kinematic crossings in terms of the same basis, augmented to accommodate the new integrals.
In practice, we use \texttt{RatRacer}~\cite{Magerya:2022hvj} to solve a linear system to determine the decomposition of every new master integral solution in the known basis of functions.
Due to the complexity and volume of genuinely distinct graphs considered in the 9 newly computed non-planar families, one might expect that the functional basis would have to be significantly enlarged.
On the contrary, we find that only 36 new functions have to be added to the original $1\,282$, for a total of $1\,318$ that capture any amplitude with this kinematics at three loops.
Clearly the three-loop problem is significantly more involved than the two-loop integrals, yet only $118$ of the $1\,318$ functions involve letters beyond the two-loop alphabet, while the remaining $1\,200$ contain only two-loop letters, see~\cref{tab:new-trident-functions}.

The 36 additional functions are of two kinds. First, there are 25 functions involving the square root letters $l_7,l_8$ that stem from the new families including $H$. The leading-colour families did feature the square root, but here it appears in one new function at weight 3, which results in a small number of higher-weight objects emanating from the iterated integration over this kernel.
Secondly, there are 11 new functions involving only the two-loop letters $l_1-l_6$, representing combinations of ordinary iterated integrals with two-loop letters at weight 6 that, by chance, did not arise at leading colour.
Note that if we had built a different canonical basis for this scattering problem, we would still arrive at the same number of minimal transcendental functions at this truncation order in $\epsilon$, because a canonical basis for every sector is unique up to $\mathbb{Q}$-linear rotations~\cite{Meyer:2017joq}.

It is striking that the letter sequences within the iterated integral tokens, and the linear combinations in which the tokens appear, are so few.
In fact, even though in $d$ dimensions 22 families contain distinct graphs and integrals, the function space up to weight 6, and thus up to the finite part of any dimensionally regulated amplitude, is spanned by significantly fewer integrals.
In other words, with the benefit of hindsight, one could build identical trident functions with as few as 10 families and their crossings, for example $A, E_1, E_2, B_1, F_1, I_1, I_2, L_3, M_3, H$.
It appears that despite the vast naive combinatorics of 20 letters arranged with repetition in up to 6 indices, the solutions of the \emph{individual} loop integrals are strictly confined by criteria shared among all the families, such as the non-appearance of certain thresholds and spurious singularities, which translate into highly restricted letter sequences ($43\,146$ distinct tokens in our case) and even fewer independent linear combinations thereof ($1\,318$ in our case).

In this context, it is remarkable that no new instances of iterated integrals with quadratic letters appear in the non-planar families contributing only to subleading colour layers. In other words, the physically allowed ways for these letters to enter the iterated integral letter sequences seem to be exhausted already by the fraction of families that contributed to the leading colour of $H$+jet.
As a result, the detailed analysis of the function space in~\cite{Gehrmann:2024tds} largely extends to the ensuing amplitudes in full colour, with the qualification about the square root at weight 3.

Besides a clean representation, the trident basis is advantageous for the decomposition of our integrals and the resulting amplitudes for three practical reasons:

\begin{table}[t]
\begin{center}
\begin{tabular}{c@{\qquad}rrrrr}
\toprule
weight & two-loop only & parabolic & hyperbolic & root & total \\
\midrule
$1$ & $3$     & $0$     & $0$     & $0$   & $3$ \\
$2$ & $13$    & $0$     & $0$     & $0$   & $13$ \\
$3$ & $60$    & $0$     & $0$     & $8$   & $68$ \\
$4$ & $318$   & $94$    & $0$     & $74$  & $486$ \\
$5$ & $1\,735$ & $376$  & $518$   & $332$ & $2\,961$ \\
$6$ & $9\,668$ & $1\,786$ & $2\,072$ & $972$ & $14\,498$ \\
\midrule
total & $11\,797$ & $2\,256$ & $2\,590$ & $1\,386$ & $18\,029$ \\
\bottomrule
\end{tabular}
\end{center}
\caption{The distribution of letter types by weight among the iterated integrals surviving after extrapolating information from the poles to the finite remainder of any ensuing amplitude.}
\label{tab:pole_blocking}
\end{table}

\begin{enumerate}
\item \textbf{Excluding functions present in the loop integrals from the finite part of any amplitude.}
On the amplitude level, the $\epsilon$-poles are predicted by previous loop orders, so they cannot contain anything other than the two-loop letters.
The grading by first appearance of a new letter effectively propagates this information
from the poles to the finite remainder.  
Indeed, suppose that a trident
function first contains one of the letters $l_{7},\ldots,l_{20}$ at weight
$w$.  Since the corresponding new-letter projections of the trident functions form an independent set by construction,
the absence of these letters from the poles forces this trident function's rational coefficient in any amplitude to
start at the power of $\epsilon$ that places the weight-$w$ term into the finite
part.  Consequently, all terms of higher weight in that function lie beyond
$\mathcal O(\epsilon^0)$ and can never contribute to the relevant part of any amplitude. 
Following this logic, if we truncate every
trident function after its first new-letter weight, we reduce the union of distinct
tokens from $43\,146$ to $18\,030$ and exclude $25\,116$ tokens with this simple argument.  Most
importantly, $24\,538$ of the excluded tokens contain a new letter: only
$6\,232$ of the original $30\,770$ such tokens survive the pole constraint, and they are qualitatively simpler than the full set.
In particular, among the surviving iterated-integral tokens, new letters of different classes never mix within one iterated integral. Quadratic letters appear at most once, while square-root letters can appear as neighbouring pairs.
Every parabolic letter occupies effective transcendental-weight slot~4, as defined in detail in~\cite{Gehrmann:2024tds}, and every hyperbolic letter occupies effective transcendental-weight slot~5.
The surviving tokens are distributed by weight and letter class as per~\cref{tab:pole_blocking}.

\item \textbf{Efficient evaluation of rational coefficients in amplitudes.} 
In an amplitude, the transcendental objects (whether master integrals or trident functions) will be multiplied by rational functions in the kinematic invariants, determined largely by a costly IBP reduction.
If we allow for cancellations between the basis elements, the functions can be arbitrarily complicated  (e.g.\ have high numerator degree).
In this basis, we have prevented most types of cancellations, and we only need to determine (which with finite-field methods amounts to reconstruction in two kinematic variables) at most 7 $\epsilon$-orders of $1\,318$ IBP coefficients.
On top of this, as discussed above, several coefficients in the $\epsilon$ expansions must vanish due to the pole argument.
Other iterated integrals are not outright excluded but the coefficients of trident functions containing them are related due to the knowledge of the poles of the amplitude, further reducing the number of coefficients to be determined.
    
\item \textbf{Complicated transcendental objects confined to a small subset.}
For storage, evaluation, analytic continuation, it is advantageous to confine the complexity in the form of quadratic and especially square root letters in a minimal set of objects.
Consider that there are $43\,146$ unique iterated integral tokens across all the integral solutions, of which $11\,300$ contain the root.
However, they fall into only 52 relevant trident functions containing the root. 
\end{enumerate}

\section{Numerical evaluation}
\label{sec:numerics}
While the iterated-integral representation constructed in \cref{sec:results}  provides a formal analytic solution of
the differential equations, its use in phenomenological applications
requires an efficient method for numerically evaluating the iterated
integrals at arbitrary kinematic points. Upon parametrising the path $\gamma$ by $t \in [t_0, t]$ with
$\gamma(t_0) = (0,0)$, \cref{eq:chen} becomes
\begin{equation}
  \int_{\gamma}\omega_1\cdots\omega_n
  = \int_{t_0}^{t}\!dt_n\,f_n(t_n)
    \int_{t_0}^{t_n}\!dt_{n-1}\,f_{n-1}(t_{n-1})
    \cdots
    \int_{t_0}^{t_2}\!dt_1\,f_1(t_1),
  \label{eq:nested}
\end{equation}
where $f_i(t_i)$ denotes the pull-back $\gamma^{*}\omega_i = dt_i\,f_i(t_i)$.
Through two loops, and for the planar topologies at three loops, the
alphabet linearises along any rectangular path, so a representation in
terms of two-dimensional harmonic polylogarithms (2dHPLs) is readily
available. The simultaneous presence of quadratic and algebraic letters
obstructs such a representation for the non-planar topologies at three
loops. In ref.~\cite{Guan:2026igf}, a generalised-polylogarithm solution
was obtained for the subset of iterated integrals entering the finite
remainder of the $H$+jet amplitude, where
non-trivial constraints limit the admissible letter sequences. At the level of individual master integrals
a considerably larger set of iterated integrals is required, making the
construction of a global polylogarithm representation more
involved.

In this section we discuss two strategies for evaluating the trident functions, both starting from the canonical systems in~\cref{eq:cde} and the boundary conditions of~\cref{sec:boundary-conditions}. Both are well established in the literature, but they have different strengths.

The first strategy builds local generalised series expansions of the solution, which are analytic and systematically improvable (see \cref{sec:series}). 
The series retain all logarithmic singularities exactly, and are therefore particularly suited for evaluation in unresolved limits. 
They also facilitate analytic continuation to the relevant scattering regions (see \cref{sec:analytic-continuation}).
The second strategy integrates the differential equations numerically along a path in complexified kinematic space, as described in \cref{sec:numerical-transport}. It requires no analytic representation of the solution, it applies equally in the decay and scattering regions, and it is well suited to probing the bulk of phase space.

The two methods are also used in combination within our approach. Numerical transport requires a boundary condition at a regular point, whereas all our boundary conditions are known on the singular lines separating the kinematic regions (see \cref{sec:boundary-conditions}). Obtaining a regular-point boundary value directly from a purely numerical evaluation (for instance with \texttt{AMFlow}~\cite{Liu:2022chg}) is computationally expensive for this problem. We therefore use the series expansions to move the boundary conditions away from the singular lines, which provides suitable starting values for the numerical transport.

We present proof-of-concept studies of both strategies and leave a complete implementation to future work.

\subsection{Numerical evaluation from accelerated series}
\label{sec:series}
\begin{figure*}[t]
  \centering
  \includegraphics[width=0.8\textwidth]{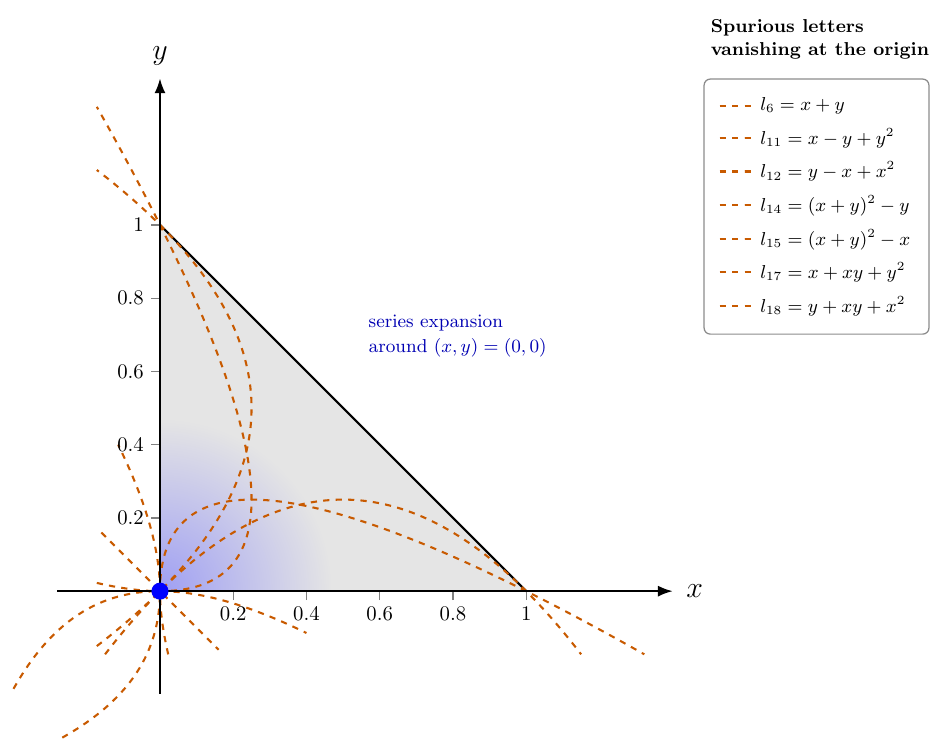}
  \caption{A representation of the Euclidean triangle. In orange, the zero loci for the letters vanishing at the origin. These correspond to spurious singularities of the differential equation system, which cancel in the physical solution.}
  \label{fig:series_origin_representation}
\end{figure*}

As a first approach, we provide a solution for all trident functions in terms of local generalised series expansions.
In this subsection, we focus on the decay region where $x,y,z >0$.
Upon expanding the solution order by order in $\epsilon$,
\begin{equation}
\boldsymbol J_{\mathcal F}(x,y;\epsilon)
=\sum_{w\geq0}\epsilon^w
\boldsymbol J_{\mathcal F}^{(w)}(x,y),
\end{equation}
the canonical system in \cref{eq:cde} implies the recursion
\begin{equation}
\diff\boldsymbol J_{\mathcal F}^{(w)}
=\diff\widetilde A_{\mathcal F}\,
\boldsymbol J_{\mathcal F}^{(w-1)}.
\label{eq:weight-recursion}
\end{equation}
We expand the connection matrix $\diff\widetilde A_{\mathcal F}(x,y)$ around a boundary point $(x_0,y_0)$ and integrate order by order. It is convenient to perform the expansion around a singular point, where physical singularities are entirely captured by corresponding towers of logarithms. As an example, let us focus on the singular expansion point $(x_0,y_0)=(0,0)$. In this region, the solution develops logarithmic singularities in either of the two variables. Consequently, the expansion of the differential equation matrix will, in general, contain simple poles.
Moreover, the differential equation system often contains unphysical singularities that cancel in the physical solutions. These spurious singularities must be resolved (blown up) before the expansion can be carried out. For the chosen point $(x_0,y_0)=(0,0)$, the only physical singularities are those associated with the letters $l_1$ and $l_2$ in \cref{eq:linletters}. Nevertheless, the spurious singular lines associated with the letters $l_6$, $l_{11}$, $l_{12}$, $l_{14}$, $l_{15}$, $l_{17}$, and $l_{18}$ all intersect at the origin, see~\cref{fig:series_origin_representation}. The change of variables
\begin{equation}
x=v\tau^2\,,\qquad y=\tau
\label{eq:origin-blowup}
\end{equation}
resolves these singularities and yields a well-ordered expansion in $\tau$ and $v$. Once the series solution has been constructed, the spurious singularities manifestly cancel, and the resulting expression can be rewritten in terms of the original variables $(x,y)$.

\begin{figure*}[t]
\centering
\includegraphics[width=0.8\textwidth]{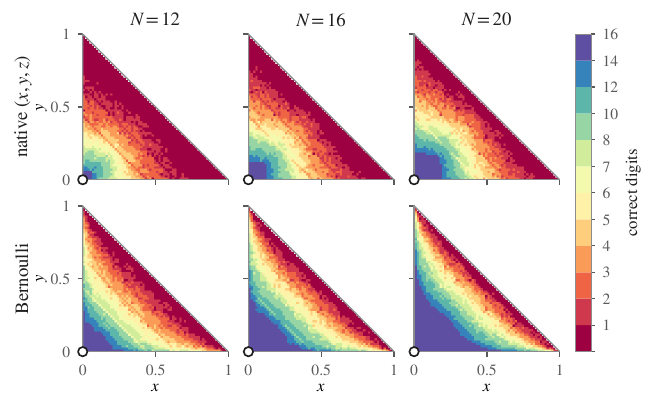}
\caption{Number of digits of agreement between the series solution centred at $(x_0,y_0)\approx(0,0)$ and the exact solution in terms of polylogarithms at weight 2 for the integrals in the $H$ family. The comparison is performed with truncation order $N=$ 12, 16 and 20. The series is generated in native $x,y,z$ and Bernoulli-like variables.
For each phase-space point, we show the number of digits corresponding to the worst agreement (fewest digits) across the 371 canonical combinations in the basis.
}
\label{fig:w2_12-16-20_c1}
\end{figure*}

The convergence of a series solution is typically limited by the nearest zero of a letter in the complexified kinematic plane. For the physical master integrals, one finds in general that not all pseudo-thresholds have an impact on the radius of convergence of the series, in particular if the original expansion point is a regular one, see for example~\cite{Pozzorini:2005ff}. Since we expand at $(0,0)$, which is a singular point, all pseudo-thresholds are expected to reduce the radius of convergence. 
Interestingly, as already observed in various calculations~\cite{Pozzorini:2005ff,Gehrmann:2001jv,Becchetti:2025rrz,Coro:2025vgn}, even in these cases the convergence of the series can be substantially improved by a Bernoulli-like change of variables which displaces the singularity nearest to the expansion point. In particular, if the limiting singularity in a local coordinate $\xi$ is located at $\xi_\star$, we introduce the new coordinate $\chi$ where
\begin{equation}
\chi=-\log\left(1-\frac{\xi}{\xi_\star}\right),
\qquad
\xi=\xi_\star\left(1-e^{-\chi}\right),
\label{eq:bernoulli-like-variable}
\end{equation}
which maps the singularity to infinity. In two variables, the transformation is applied successively. The final transformation is given by
\begin{align}
s_x = -\log(1-x), \qquad s_y = -\log(1-y).
\label{eq:bernoulli-xy}
\end{align}
This is the multivariate analogue of the convergence acceleration used for two-dimensional harmonic polylogarithms in ref.~\cite{Gehrmann:2001jv}.
We illustrate this behaviour in~\cref{fig:w2_12-16-20_c1}. Both series exhibit improved convergence as the truncation order is increased, with faster convergence when Bernoulli-like variables are used.

\begin{figure*}[t]
\centering
\includegraphics[width=0.8\textwidth]{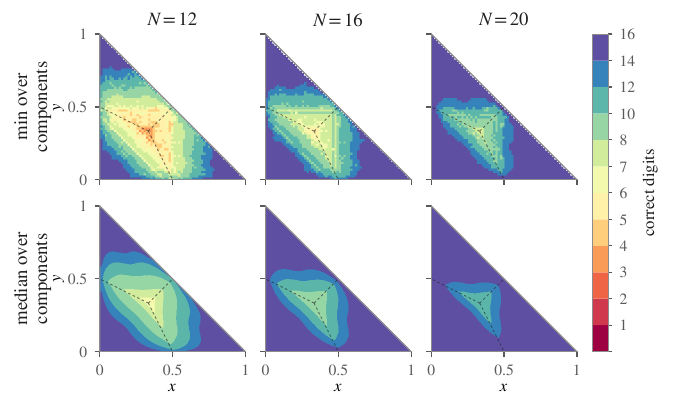}
\caption{Number of digits of agreement between the series solution truncated at orders 12, 16, and 20 in the Bernoulli-like variables $s_x,s_y$ and the exact solution in terms of polylogarithms at weight 2 for the integrals in the $H$ family. For each phase-space point, the choice of the expansion centre is made based on which of the three variables $x,y,z$ is the largest.  
Two measures of the number of digits are shown: \textit{min}, corresponding to the worst agreement (fewest digits), and \textit{median}, corresponding to the median agreement across the 371 canonical combinations in the basis.}
\label{fig:w2_12-16-20_cov}
\end{figure*}

Nonetheless, no single expansion is expected to be optimal over the entire decay triangle. This can be understood by inspecting the physical singularities of our integrals. In particular, the line $y=1-x$ in~\cref{fig:series_origin_representation} corresponds to the singular limit $z=0$. However, any expansion around $(0,0)$ in the coordinates $x$ and $y$ treats $\log(z)$ as regular away from the boundary and therefore cannot capture its singular behaviour as $z\to0$. Moreover, the two-variable Bernoulli-like change of variables in~\cref{eq:bernoulli-xy} does not map the entire singular line $y=1-x$ to infinity.

\begin{figure*}[t]
\centering

\includegraphics[width=0.8\textwidth]{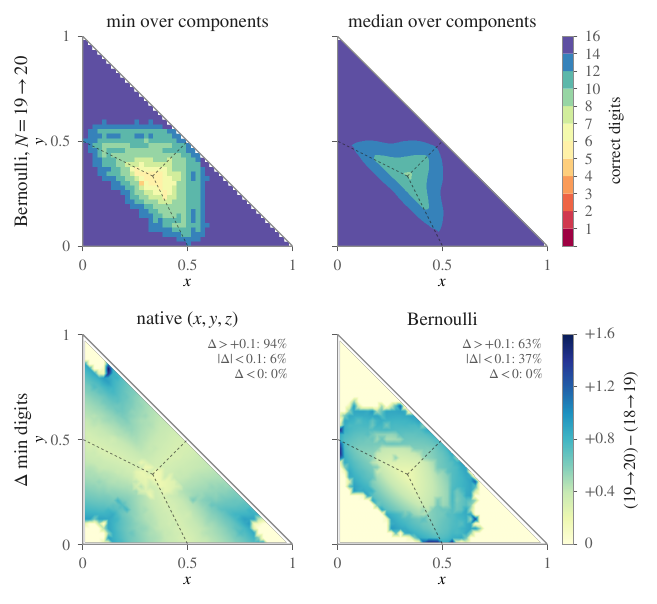}
\caption{
Partial-sum test for the integrals in the H family through weight 6. 
For each phase-space point, one of the three corners of the triangle is chosen as the expansion point based on which of the variables $x$, $y$, $z$ has the largest value. 
\emph{Top row:} number of digits of agreement between the series in the Bernoulli-like variables $s_x$, $s_y$ truncated at orders 19 and 20. The two panels display the minimum (left) and median (right) agreement across the 371 master integrals. 
\emph{Bottom row:} difference in the minimum agreement across the integral basis when comparing orders $(18, 19)$ and $(19, 20)$, using the native variables $x$, $y$, $z$ (left) and the Bernoulli-like variables (right). 
A positive value means that the agreement improves with truncation order; no phase-space point shows a decrease. The percentages in the legend give the distribution of the differences in minimum agreement between the pairs of orders across the phase-space points.}
\label{fig:w6_self_conv}
\end{figure*}

We address this issue by constructing series expansions around the three corners of the kinematic triangle, corresponding to the limits in which two of the three variables $x,y,z$ vanish, while the third variable approaches $1$ due to momentum conservation. At each phase-space point, we select the expansion associated with the corner opposite the largest variable. For example, if $z$ is the largest coordinate, the corner $x\approx0$, $y\approx0$ is selected. We illustrate this strategy for family $H$ in~\cref{fig:w2_12-16-20_cov}. Up to transcendental weight 2, only the linear letters from~\cref{eq:linletters} appear,
which can easily be converted to classical polylogarithms and evaluated to arbitrary precision as a reference.
Truncating our series representation at order 20, an agreement of $6$--$8$ digits with the reference is achieved throughout the entire kinematic region and across the full canonical basis. Finally, in~\cref{fig:w6_self_conv} we perform a partial-sum test for family $H$ through transcendental weight 6. Note that the series solutions for crossings can easily be built out of the uncrossed series in the three corners by permuting the kinematic variables. 

Several recent multiloop calculations, including some with elliptic kernels and beyond, have found that evaluating the amplitude as a whole performs significantly better than evaluating its individual master integrals with the same method. There are two main reasons for this. First, often only a subset of the canonical integrals contribute to the finite remainder. Second, these integrals enter in combinations in which the structures that are large or hard to evaluate precisely in the individual integrals cancel in certain kinematic limits. The amplitude then behaves like a much simpler function, as dictated by infrared factorisation. Performing the series expansion directly at the amplitude level for both the rational coefficients and the surviving iterated integrals exploits this and often converges faster. Our results should therefore be read as a lower bound on the achievable performance.

For the present work, we generated the series expansions in the three corners up to order 20 for all the integral families considered.
We checked the numerical values for a sample of top-sector integrals from each of the published families against numerical evaluation obtained with the packages~\texttt{AMFlow} and~\texttt{FIESTA5}.

\subsubsection{Continuation to scattering regions}
\label{sec:analytic-continuation}
The numerical results discussed above are for the decay kinematic region, 
i.e. where the off-shell particle decays into 3 massless particles. 
The $2\to2$ scattering regions relevant for the LHC can be accessed through analytic continuation of the special functions.
The procedure was described in detail for this kinematics up to two loops in~\cite{Gehrmann:2002zr,Gehrmann:2023etk}.

\begin{figure}[t!]
\centering
\includegraphics[width=0.75\textwidth]{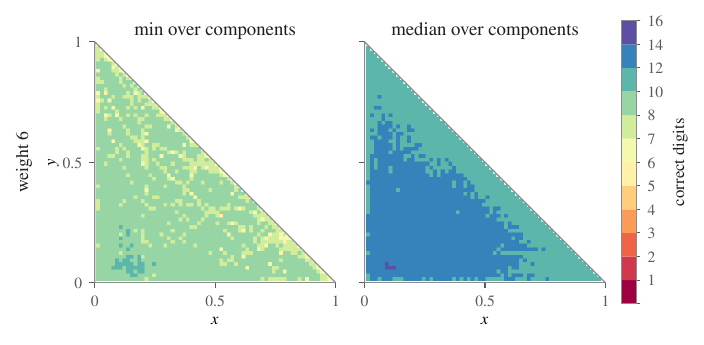}
\caption{Estimated number of correct digits for the weight-six functions of family $H$ in the decay region, using double precision arithmetic (tolerance $10^{-10}$). By ``components'' we indicate the various $\epsilon$ orders of each of the $371$ master integrals. The estimate is obtained by comparing evaluations with deformations $0.1$ and $0.4$ of the integration path, at $2016$ lattice points. \emph{Left:} the worst, \emph{right:} the median agreement over the $371$ functions.}
\label{fig:H-precision-map}
\end{figure}

For the present extension to three loops, the series construction also gives a direct implementation of analytic
continuation. 
The dimensionful invariants are first endowed with the
imaginary parts dictated by the Feynman prescription, and the ratios $x$, $y$
and $z$ are formed only afterwards.  Starting from the decay region, a path in
the complexified kinematic plane is chosen that avoids the zero loci of the
alphabet.  Along this path, the square root $\rho$ is analytically continued,
which fixes the branches of the two algebraic letters in
\cref{eq:algletters}.  Local series patches can then be matched along
the path, or constructed directly inside the desired scattering region.  As a result, this gives stable evaluations in the production region
$x>0$, $y<0$, $z<0$ without requiring a separate global polylogarithmic
representation.

\subsection{Numerical evaluation by transport of the canonical system}
\label{sec:numerical-transport}
In a second approach, we follow the strategy of~\cite{Czakon:2026tbd}, in which the canonical differential equations are used to transport the integral solutions numerically along a path in complexified kinematic space. The method also applies beyond polylogarithms, but it is simplest when the connection is written in d-log form, as in~\cref{eq:cde}. Algebraic letters pose no obstacle: since $\diff\sqrt{g}=\sqrt{g}\,\diff g/(2g)$, the square roots can be appended and transported together with the integrals as a single system of rational differential equations. Given a boundary value, the system is integrated with the \texttt{odeint} routines of the \texttt{Boost} libraries~\cite{Ahnert:2011odeint}.
We adapt the public \texttt{C++} package accompanying~\cite{Czakon:2026tbd} to evaluate the canonical integrals of family~$H$, in both the decay region and one representative scattering region. Note that this is a proof of concept: a production-grade implementation should be tailored to the amplitude rather than to the integrals individually.

The canonical system of family $H$ has dimension $371$.
We transport numerically all integrals through weight six, together with the square root $\rho$, as a single system, so that all $2226$ functions of weights one to six are obtained simultaneously at each point.
We work in double precision, set the absolute and relative local tolerance of the code to $10^{-10}$ 
and evaluate integrals in the decay region and one scattering region (we choose $x,y<0$, so that $z>1$) at $2016$ points each, uniformly distributed.\footnote{Our setup for the local tolerance means in practice that a step is accepted if the estimated local error of every component $f_i$ of the system is below $10^{-10}(1 + |f_i|)$.}
Since the code requires a boundary value at a non-singular point, 
we employ the series expansion solutions in~\cref{sec:series} to evolve our boundary from $(x_0,y_0) = (0,0)$, where the integrals are logarithmically divergent, 
to $(x_0,y_0)=(0.1,\,0.07)$ in the decay region and
$(x_0,y_0)=(-0.1,\,-0.07)$, i.e.\ $z_0=1.17$, in the scattering region.
Using our Bernoulli-improved series, we can obtain the values of all $371$ canonical integrals
at these points through weight six to about $35$ correct digits.\footnote{The number of correct digits of the results obtained from the series expansions is estimated, as described in \cref{sec:series} above, by comparing partial sums.} 

\begin{table}[t!]
\begin{center}
\begin{tabular}{l@{\qquad}cc}
\toprule
 & decay region & scattering region \\
\midrule
median time per point [s] & $0.14$ & $0.14$ \\
mean time per point [s] & $0.15$ & $0.16$ \\
maximum [s] & $0.35$ & $0.47$ \\
\midrule
core-hours for $2016$ points & $0.08$ & $0.09$ \\
\bottomrule
\end{tabular}
\end{center}
\caption{Time to evaluate all $2226$ functions of weights one to six of family $H$, together with the square root, at one point, in double precision (tolerance $10^{-10}$) on a single state-of-the-art core. Time per point: three processes in parallel, $2016$ points per region.}
\label{tab:H-timing}
\end{table}

From these two boundary points, 
the solution is transported along a straight line to the target point, which is made possible by the convexity of the kinematic regions. 
We estimate the accuracy  by comparing two evaluations whose paths to the same endpoint are 
deformed into the complex plane 
by different amounts. For this purpose, 
the integration path from the boundary point $\vec{\xi}_0 = (x_0,y_0)$ to the target point $\vec{\xi}_1=(x_1,y_1)$ is parametrised as 
\begin{align}
    \vec{\zeta}(t)= \vec{\xi}_0+\bigl(1+4i\,\eta\,(1-t)\bigr)\,t\,(\vec{\xi}_1-\vec{\xi}_0) \nonumber
\end{align}
for $t\in[0,1]$, so that its imaginary part vanishes at both ends
and reaches $\eta\,(\vec{\xi}_1-\vec{\xi}_0)$ at the midpoint.
The two deformation paths correspond to 
deformation parameters $\eta=0.1$ and $\eta=0.4$.\footnote{Note that, in general, one could deform the path asymmetrically in the $x$ and $y$ directions. For this exploratory application, we observe a symmetric deformation to be already sufficient.} 
We checked this estimate against a double-double evaluation and found that the estimate is conservative, underestimating the true accuracy by $0.2$ digits on average.
Across all weights, 
the median accuracy, taken over all functions of a given weight and all points in the region, is about 11 digits. Moreover, at $99\%$ of the lattice points, all 2226 functions are correct to at least 5 digits, and at $94\%$ to at least 6. 
A summary of the precision at weight six is provided 
in~\cref{fig:H-precision-map}, while 
 evaluation times are provided in~\cref{tab:H-timing}.
 The same strategy applies, with minor modifications, directly to the trident functions, since weight by weight they also satisfy inhomogeneous systems of rational differential equations.

\section{Conclusion}
\label{sec:concl}

We have completed the calculation of the massless three-loop four-point master integrals with one off-shell external leg. For all 18 irreducible integral families, including the 9 previously unpublished non-planar families, and for 4 reducible families spanning the remaining integrals, we constructed canonical bases directly from integrands with logarithmic singularities and constant leading singularities. The resulting differential equations close on a common 20-letter alphabet, and their boundary conditions and solutions were determined through transcendental weight six. This includes the 371 master integrals of family $H$ and its 19-dimensional top sector. Despite the complexity of the individual families, all solutions and their kinematic crossings span a compact space of $1\,318$ independent functions, only 36 more than were required for the subset of families contributing to the $H$+jet amplitude in the leading-colour approximation. 

The iterated-integral solutions provide an analytic description of the complete function space, and we have discussed two routes for their numerical evaluation: accelerated local series expansions and numerical transport. While the former offer a stable and systematically improvable route to numerical evaluation and analytic continuation in physical scattering regions, the latter is directly applicable as long as a boundary condition in the appropriate kinematic region is known. In particular, our results supply the missing integral input for three-loop vector-boson-plus-jet production beyond the planar approximation and open the way to full-colour predictions for \(V+\)jet production at third order in the strong coupling.

\section*{Acknowledgements}
We thank Johannes Henn and William Torres Bobadilla for collaboration in the early stages of this project and for useful comments on the paper.
We have used AI agents, including OpenAI Codex 5.6, for code implementation and in the editing process of this manuscript.
This work has been supported by the 
Excellence Cluster ORIGINS funded by the Deutsche
Forschungsgemeinschaft (DFG, German Research Foundation) under Germany’s Excellence Strategy – EXC-2094-390783311, by the European Research Council (ERC) under grant agreements 949279 (ERC
Starting Grant HighPHun) and 101019620 (ERC Advanced Grant TOPUP),
by the Swiss National Science Foundation (SNF) under contract 240015, by the UKRI Frontier Research Grant, underwriting the ERC Consolidator Grant precSM (UKRI946), 
and by the Science and Technology Facilities Council (STFC) under grant ST/X000761/1.

\appendix

\section{Improved Baikov representation for $I_3$}
\label{app:improved-baikov-I3}
In this appendix, we describe how the improved Baikov representation can be used to study the three-loop integral family $I_3$ shown in \cref{fig:one-mass-integral-families}. 
The main advantage of this representation is that suitable one-loop box building blocks provide variables in which part of the four-dimensional integration measure naturally takes a d-log form. 
This simplifies the computation of generalised cuts and provides
a useful starting point for constructing canonical-basis candidates.

\subsection{The one-loop building block}

We begin by recalling the d-log representation of the one-loop box integral. 
Consider the four propagators
\begin{equation}
\begin{aligned}
D_1 &= k^2,\\
D_2 &= (k+p_1)^2,\\
D_3 &= (k+p_1+p_2)^2,\\
D_4 &= (k-p_3)^2,
\end{aligned}
\end{equation}
where $p_1^2=p_2^2=p_3^2=0$ and
$p_4=-p_1-p_2-p_3$.
The quadruple-cut equations,
\begin{equation}
D_1=D_2=D_3=D_4=0,
\end{equation}
have two solutions which we denote by $k_*^\pm$, 
with both solutions giving valid changes of variables.
For either choice of solution, introduce four new variables
\begin{equation}
\alpha_i =
\frac{D_i}{(k-k_*^\pm)^2},
\qquad i=1,\ldots,4.
\label{eq:I3-alpha-def}
\end{equation}
As shown in ref.~\cite{Henn:2019swt},
the normalised box integrand admits the representation
\begin{equation}
\frac{s t\,\diff^4 k}{D_1D_2D_3D_4}
=
\pm
\diff \log\alpha_1\wedge
\diff \log\alpha_2\wedge
\diff \log\alpha_3\wedge
\diff \log\alpha_4,
\label{eq:I3-box-dlog}
\end{equation}
where $s=(p_1+p_2)^2$ and $t=(p_1+p_3)^2$.

The important feature of this transformation is that each propagator is associated with a single variable.
On the appropriate coordinate patch, the cut of a propagator is therefore implemented simply by taking the residue at the corresponding $\alpha_i=0$.
The two solutions of the quadruple cut describe the two possible branches, which can be analysed by choosing the corresponding coordinate patches.

\subsection{Application to the $I_3$ family}
\label{subsec:I3-improved-baikov}

The top-sector canonical candidates for the $I_3$ family were
originally constructed using \texttt{DlogBasis} together with the
improved Baikov representation described below. 
The latter provides a particularly simple parametrisation of the leading-singularity problem. 
With the improvements to \texttt{DlogBasis} described in~\cref{sec:dlog-strategy}, the candidates can now also be obtained directly from the four-dimensional loop-momentum parametrisation. 
We nevertheless describe the improved Baikov
construction here, since it makes the d-log structure of the
$I_3$ integrand particularly transparent.

\begin{figure}[h!]
    \centering
    \includegraphics[width=0.9\textwidth]{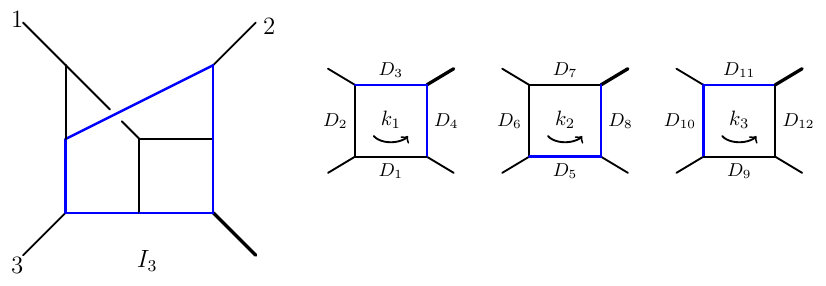}
    \caption{
    The non-planar family $I_3$ and the three one-loop box
    building blocks used in the improved Baikov representation.
    The propagators that are directly identified between the
    $I_3$ topology and the corresponding boxes are highlighted
    in blue.
    }
    \label{fig:I3-box-decomposition}
\end{figure}

The main observation is that the topology contains three natural
one-loop box building blocks, associated with the loop momenta
$k_1$, $k_2$, and $k_3$. 
The four propagators of the three boxes are labelled consecutively as $D_1,\ldots,D_4$, $D_5,\ldots,D_8$, and $D_9,\ldots,D_{12}$, respectively.
As illustrated in \cref{fig:I3-box-decomposition}, six of these box propagators can be directly identified with propagators of the $I_3$ topology, with two inherited from each box.

For each box, the remaining loop momenta are kept fixed and the corresponding quadruple-cut equations are solved. Denoting the two solutions for the box associated with $k_r$ by $k_{r,*}^{\pm}$, we introduce the improved Baikov variables
\begin{equation}
    \alpha_i^{(r)}
    =
    \frac{D_i}{(k_r-k_{r,*}^{\pm})^2},
    \qquad
    i=4(r-1)+1,\ldots,4r,
    \qquad r=1,2,3.
\end{equation}
In the following, we suppress the superscript $(r)$ and label the twelve variables simply by $\alpha_1,\ldots,\alpha_{12}$.

Taken together, the three box transformations provide a local twelve-dimensional change of variables for the three loop momenta.
The six box propagators $\{D_3,D_4,D_5,D_8,D_{10},D_{11}\}$ coincide with propagators of the $I_3$ topology and therefore supply six logarithmic directions directly. 
Suppressing overall normalisation factors and the dependence on the choice of quadruple-cut branch, the four-dimensional integrand can be written schematically as
\begin{align}
\Omega_{I_3}^{(4)}
\sim\;&
\diff\log\alpha_3
\wedge \diff \log\alpha_4
\wedge \diff \log\alpha_5
\wedge \diff \log\alpha_8
\wedge \diff \log\alpha_{10}
\wedge \diff \log\alpha_{11}
\nonumber\\
&\wedge\,
\diff \alpha_1\wedge \diff \alpha_2
\wedge \diff \alpha_6\wedge \diff \alpha_7
\wedge \diff \alpha_9\wedge \diff \alpha_{12}\,
R_{I_3}(D_1,\ldots,D_{12}) .
\label{eq:I3-improved-baikov}
\end{align}
Here the $D_i$ are understood as functions of the improved Baikov
variables, while the remaining dependence is encoded in the rational
function $R_{I_3}$.

The parametrisation has two main advantages. First, cuts of the
highlighted propagators are immediate: by construction,
\begin{equation}
    D_i=0
    \qquad\Longleftrightarrow\qquad
    \alpha_i=0 ,
\end{equation}
so in practice the desired cuts are imposed simply by setting the
corresponding $D_i$ to zero.

Second, after extracting the three box building blocks,
$R_{I_3}(D_1,\ldots,D_{12})$ is substantially simpler than the
Baikov polynomial of the conventional Baikov representation.
The leading-singularity analysis is therefore reduced to a much
simpler rational problem, with six logarithmic directions already
manifest.

The improved Baikov representation therefore provides an efficient
parametrisation of the leading-singularity problem for $I_3$.
Combined with \texttt{DlogBasis}, it was used to construct the
top-sector canonical candidates and considerably simplify the
remaining residue calculation.

%---------bibliography if using bibtex
\bibliographystyle{JHEP} % doi-Link, when doi  = {10.1039/C3CC46767H} in .bib file
\bibliography{biblio} % with XeLaTex: needs the same name as main texfile

@article{Pozzorini:2005ff,
    author = "Pozzorini, S. and Remiddi, E.",
    title = "{Precise numerical evaluation of the two loop sunrise graph master integrals in the equal mass case}",
    eprint = "hep-ph/0505041",
    archivePrefix = "arXiv",
    reportNumber = "TTP05-06, SFB-CPP-05-14",
    doi = "10.1016/j.cpc.2006.05.005",
    journal = "Comput. Phys. Commun.",
    volume = "175",
    pages = "381--387",
    year = "2006"
}

@article{Gehrmann:2001jv,
    author = "Gehrmann, T. and Remiddi, E.",
    title = "{Numerical evaluation of two-dimensional harmonic polylogarithms}",
    eprint = "hep-ph/0111255",
    archivePrefix = "arXiv",
    reportNumber = "CERN-TH-2001-326",
    doi = "10.1016/S0010-4655(02)00139-X",
    journal = "Comput. Phys. Commun.",
    volume = "144",
    pages = "200--223",
    year = "2002"
}

@article{Coro:2025vgn,
    author = "Coro, Federico and Nega, Christoph and Tancredi, Lorenzo and Wagner, Fabian J.",
    title = "{Analytic two-loop amplitudes for di-jet and {\ensuremath{\gamma}}+jet production mediated by a heavy-quark loop}",
    eprint = "2509.15315",
    archivePrefix = "arXiv",
    primaryClass = "hep-ph",
    reportNumber = "TUM-HEP 1570/25",
    doi = "10.1007/JHEP01(2026)090",
    journal = "JHEP",
    volume = "01",
    pages = "090",
    year = "2026"
}

@article{Becchetti:2025rrz,
    author = "Becchetti, Matteo and Coro, Federico and Nega, Christoph and Tancredi, Lorenzo and Wagner, Fabian J.",
    title = "{Analytic two-loop amplitudes for $ q\overline{q}\to \gamma \gamma $ and gg {\textrightarrow} {\ensuremath{\gamma}}{\ensuremath{\gamma}} mediated by a heavy-quark loop}",
    eprint = "2502.00118",
    archivePrefix = "arXiv",
    primaryClass = "hep-ph",
    reportNumber = "TUM-HEP 1554/25",
    doi = "10.1007/JHEP06(2025)033",
    journal = "JHEP",
    volume = "06",
    pages = "033",
    year = "2025"
}

@article{Gehrmann:2011ab,
    author = "Gehrmann, Thomas and Tancredi, Lorenzo",
    title = "{Two-loop QCD helicity amplitudes for $q\bar q \to W^\pm \gamma$ and $q\bar q \to Z^0 \gamma$}",
    eprint = "1112.1531",
    archivePrefix = "arXiv",
    primaryClass = "hep-ph",
    reportNumber = "ZU-TH-26-11, ZU-TH 26/11",
    doi = "10.1007/JHEP02(2012)004",
    journal = "JHEP",
    volume = "02",
    pages = "004",
    year = "2012"
}

@article{Gehrmann:2002zr,
    author = "Gehrmann, T. and Remiddi, E.",
    title = "{Analytic continuation of massless two loop four point functions}",
    eprint = "hep-ph/0207020",
    archivePrefix = "arXiv",
    reportNumber = "CERN-TH-2002-145",
    doi = "10.1016/S0550-3213(02)00569-2",
    journal = "Nucl. Phys. B",
    volume = "640",
    pages = "379--411",
    year = "2002"
}

@article{Henn:2013pwa,
    author = "Henn, Johannes M.",
    title = "{Multiloop integrals in dimensional regularization made simple}",
    eprint = "1304.1806",
    archivePrefix = "arXiv",
    primaryClass = "hep-th",
    doi = "10.1103/PhysRevLett.110.251601",
    journal = "Phys. Rev. Lett.",
    volume = "110",
    pages = "251601",
    year = "2013"
}

@article{Meyer:2017joq,
    author = "Meyer, Christoph",
    title = "{Algorithmic transformation of multi-loop master integrals to a canonical basis with CANONICA}",
    eprint = "1705.06252",
    archivePrefix = "arXiv",
    primaryClass = "hep-ph",
    reportNumber = "HU-EP-17-10",
    doi = "10.1016/j.cpc.2017.09.014",
    journal = "Comput. Phys. Commun.",
    volume = "222",
    pages = "295--312",
    year = "2018"
}

@article{Prausa:2017ltv,
    author = "Prausa, Mario",
    title = "{epsilon: A tool to find a canonical basis of master integrals}",
    eprint = "1701.00725",
    archivePrefix = "arXiv",
    primaryClass = "hep-ph",
    reportNumber = "TTK-17-01",
    doi = "10.1016/j.cpc.2017.05.026",
    journal = "Comput. Phys. Commun.",
    volume = "219",
    pages = "361--376",
    year = "2017"
}

@article{Lee:2020zfb,
    author = "Lee, Roman N.",
    title = "{Libra: A package for transformation of differential systems for multiloop integrals}",
    eprint = "2012.00279",
    archivePrefix = "arXiv",
    primaryClass = "hep-ph",
    doi = "10.1016/j.cpc.2021.108058",
    journal = "Comput. Phys. Commun.",
    volume = "267",
    pages = "108058",
    year = "2021"
}

@article{DiVita:2014pza,
    author = "Di Vita, Stefano and Mastrolia, Pierpaolo and Schubert, Ulrich and Yundin, Valery",
    title = "{Three-loop master integrals for ladder-box diagrams with one massive leg}",
    eprint = "1408.3107",
    archivePrefix = "arXiv",
    primaryClass = "hep-ph",
    doi = "10.1007/JHEP09(2014)148",
    journal = "JHEP",
    volume = "09",
    pages = "148",
    year = "2014"
}

@article{Canko:2021xmn,
    author = "Canko, Dhimiter D. and Syrrakos, Nikolaos",
    title = "{Planar three-loop master integrals for 2 \textrightarrow{} 2 processes with one external massive particle}",
    eprint = "2112.14275",
    archivePrefix = "arXiv",
    primaryClass = "hep-ph",
    reportNumber = "TUM-HEP-1385/21",
    doi = "10.1007/JHEP04(2022)134",
    journal = "JHEP",
    volume = "04",
    pages = "134",
    year = "2022"
}

@article{Henn:2023vbd,
    author = "Henn, Johannes M. and Lim, Jungwon and Torres Bobadilla, William J.",
    title = "{First look at the evaluation of three-loop non-planar Feynman diagrams for Higgs plus jet production}",
    eprint = "2302.12776",
    archivePrefix = "arXiv",
    primaryClass = "hep-th",
    reportNumber = "MPP-2023-35",
    doi = "10.1007/JHEP05(2023)026",
    journal = "JHEP",
    volume = "05",
    pages = "026",
    year = "2023"
}

@article{Chicherin:2020umh,
    author = "Chicherin, Dimitry and Henn, Johannes M. and Papathanasiou, Georgios",
    title = "{Cluster algebras for Feynman integrals}",
    eprint = "2012.12285",
    archivePrefix = "arXiv",
    primaryClass = "hep-th",
    reportNumber = "DESY-20-204, MPP-2020-216",
    doi = "10.1103/PhysRevLett.126.091603",
    journal = "Phys. Rev. Lett.",
    volume = "126",
    number = "9",
    pages = "091603",
    year = "2021"
}

@article{Syrrakos:2023mor,
    author = "Syrrakos, Nikolaos and Canko, Dhimiter D.",
    title = "{Three-loop master integrals for H+jet production at N3LO: Towards the non-planar topologies}",
    eprint = "2307.08432",
    archivePrefix = "arXiv",
    primaryClass = "hep-ph",
    reportNumber = "TUM-HEP-1463/23",
    doi = "10.22323/1.432.0044",
    journal = "PoS",
    volume = "RADCOR2023",
    pages = "044",
    year = "2024"
}

@article{Gehrmann:2023jyv,
    author = "Gehrmann, Thomas and Jakub\v{c}\'\i{}k, Petr and Mella, Cesare Carlo and Syrrakos, Nikolaos and Tancredi, Lorenzo",
    title = "{Planar three-loop QCD helicity amplitudes for V+jet production at hadron colliders}",
    eprint = "2307.15405",
    archivePrefix = "arXiv",
    primaryClass = "hep-ph",
    reportNumber = "ZU-TH 38/23, TUM-HEP 1467/23",
    doi = "10.1016/j.physletb.2023.138369",
    journal = "Phys. Lett. B",
    volume = "848",
    pages = "138369",
    year = "2024"
}

@article{Gehrmann:2000zt,
    author = "Gehrmann, T. and Remiddi, E.",
    title = "{Two loop master integrals for $\gamma^*\to$ 3 jets: The Planar topologies}",
    eprint = "hep-ph/0008287",
    archivePrefix = "arXiv",
    reportNumber = "KA-TTP-00-20",
    doi = "10.1016/S0550-3213(01)00057-8",
    journal = "Nucl. Phys. B",
    volume = "601",
    pages = "248--286",
    year = "2001"
}

@article{Gehrmann:2001ck,
    author = "Gehrmann, T. and Remiddi, E.",
    title = "{Two loop master integrals for $\gamma^*\to$  3 jets: The Nonplanar topologies}",
    eprint = "hep-ph/0101124",
    archivePrefix = "arXiv",
    reportNumber = "CERN-TH-2001-005",
    doi = "10.1016/S0550-3213(01)00074-8",
    journal = "Nucl. Phys. B",
    volume = "601",
    pages = "287--317",
    year = "2001"
}

@article{Gehrmann:2011aa,
    author = "Gehrmann, T. and Jaquier, M. and Glover, E. W. N. and Koukoutsakis, A.",
    title = "{Two-Loop QCD Corrections to the Helicity Amplitudes for $H \to$ 3 partons}",
    eprint = "1112.3554",
    archivePrefix = "arXiv",
    primaryClass = "hep-ph",
    reportNumber = "IPPP-11-77, ZU-TH-24-11",
    doi = "10.1007/JHEP02(2012)056",
    journal = "JHEP",
    volume = "02",
    pages = "056",
    year = "2012"
}

@article{Gehrmann:2023etk,
    author = "Gehrmann, Thomas and Jakub\v{c}\'\i{}k, Petr and Mella, Cesare Carlo and Syrrakos, Nikolaos and Tancredi, Lorenzo",
    title = "{Two-loop helicity amplitudes for $H+$jet production to higher orders in the dimensional regulator}",
    eprint = "2301.10849",
    archivePrefix = "arXiv",
    primaryClass = "hep-ph",
    reportNumber = "ZU-TH 07/23, TUM-HEP 1451/23",
    doi = "10.1007/JHEP04(2023)016",
    journal = "JHEP",
    volume = "04",
    pages = "016",
    year = "2023"
}

@article{Dixon:2020bbt,
    author = "Dixon, Lance J. and McLeod, Andrew J. and Wilhelm, Matthias",
    title = "{A Three-Point Form Factor Through Five Loops}",
    eprint = "2012.12286",
    archivePrefix = "arXiv",
    primaryClass = "hep-th",
    reportNumber = "SLAC-PUB-17581",
    doi = "10.1007/JHEP04(2021)147",
    journal = "JHEP",
    volume = "04",
    pages = "147",
    year = "2021"
}

@article{Kotikov:1991pm,
    author = "Kotikov, A. V.",
    title = "{Differential equation method: The Calculation of N point Feynman diagrams}",
    doi = "10.1016/0370-2693(91)90536-Y",
    journal = "Phys. Lett. B",
    volume = "267",
    pages = "123--127",
    year = "1991",
    note = "[Erratum: Phys.Lett.B 295, 409--409 (1992)]"
}

@article{Gehrmann:1999as,
    author = "Gehrmann, T. and Remiddi, E.",
    title = "{Differential equations for two-loop four-point functions}",
    eprint = "hep-ph/9912329",
    archivePrefix = "arXiv",
    reportNumber = "TTP-99-49",
    doi = "10.1016/S0550-3213(00)00223-6",
    journal = "Nucl. Phys. B",
    volume = "580",
    pages = "485--518",
    year = "2000"
}

@article{Remiddi:1997ny,
    author = "Remiddi, Ettore",
    title = "{Differential equations for Feynman graph amplitudes}",
    eprint = "hep-th/9711188",
    archivePrefix = "arXiv",
    reportNumber = "DFUB-97-15, DFUB 97-15",
    doi = "10.1007/BF03185566",
    journal = "Nuovo Cim. A",
    volume = "110",
    pages = "1435--1452",
    year = "1997"
}

@article{Chetyrkin:1981qh,
    author = "Chetyrkin, K. G. and Tkachov, F. V.",
    title = "{Integration by parts: The algorithm to calculate $\beta$-functions in 4 loops}",
    doi = "10.1016/0550-3213(81)90199-1",
    journal = "Nucl. Phys. B",
    volume = "192",
    pages = "159--204",
    year = "1981"
}

@article{Laporta:2000dsw,
    author = "Laporta, S.",
    title = "{High-precision calculation of multiloop Feynman integrals by difference equations}",
    eprint = "hep-ph/0102033",
    archivePrefix = "arXiv",
    doi = "10.1142/S0217751X00002159",
    journal = "Int. J. Mod. Phys. A",
    volume = "15",
    pages = "5087--5159",
    year = "2000"
}

@article{vonManteuffel:2014ixa,
    author = "von Manteuffel, Andreas and Schabinger, Robert M.",
    title = "{A novel approach to integration by parts reduction}",
    eprint = "1406.4513",
    archivePrefix = "arXiv",
    primaryClass = "hep-ph",
    reportNumber = "MITP-14-009",
    doi = "10.1016/j.physletb.2015.03.029",
    journal = "Phys. Lett. B",
    volume = "744",
    pages = "101--104",
    year = "2015"
}

@article{Peraro:2016wsq,
    author = "Peraro, Tiziano",
    title = "{Scattering amplitudes over finite fields and multivariate functional reconstruction}",
    eprint = "1608.01902",
    archivePrefix = "arXiv",
    primaryClass = "hep-ph",
    reportNumber = "EDINBURGH-2016-14",
    doi = "10.1007/JHEP12(2016)030",
    journal = "JHEP",
    volume = "12",
    pages = "030",
    year = "2016"
}

@article{Lange:2025fba,
    author = "Lange, Fabian and Usovitsch, Johann and Wu, Zihao",
    title = "{Kira 3: integral reduction with efficient seeding and optimized equation selection}",
    eprint = "2505.20197",
    archivePrefix = "arXiv",
    primaryClass = "hep-ph",
    reportNumber = "ZU-TH 39/25, HU-EP-25/17-RTG",
    doi = "10.1016/j.cpc.2025.109999",
    journal = "Comput. Phys. Commun.",
    volume = "322",
    pages = "109999",
    year = "2026"
}

@article{Boughezal:2015dra,
    author = "Boughezal, Radja and Caola, Fabrizio and Melnikov, Kirill and Petriello, Frank and Schulze, Markus",
    title = "{Higgs boson production in association with a jet at next-to-next-to-leading order}",
    eprint = "1504.07922",
    archivePrefix = "arXiv",
    primaryClass = "hep-ph",
    reportNumber = "CERN-PH-TH-2015-056, TTP15-017",
    doi = "10.1103/PhysRevLett.115.082003",
    journal = "Phys. Rev. Lett.",
    volume = "115",
    number = "8",
    pages = "082003",
    year = "2015"
}

@article{Boughezal:2015aha,
    author = "Boughezal, Radja and Focke, Christfried and Giele, Walter and Liu, Xiaohui and Petriello, Frank",
    title = "{Higgs boson production in association with a jet at NNLO using jettiness subtraction}",
    eprint = "1505.03893",
    archivePrefix = "arXiv",
    primaryClass = "hep-ph",
    reportNumber = "FERMILAB-PUB-15-210-T",
    doi = "10.1016/j.physletb.2015.06.055",
    journal = "Phys. Lett. B",
    volume = "748",
    pages = "5--8",
    year = "2015"
}

@article{Caola:2015wna,
    author = "Caola, Fabrizio and Melnikov, Kirill and Schulze, Markus",
    title = "{Fiducial cross sections for Higgs boson production in association with a jet at next-to-next-to-leading order in QCD}",
    eprint = "1508.02684",
    archivePrefix = "arXiv",
    primaryClass = "hep-ph",
    reportNumber = "CERN-PH-TH-2015-192, TTP15-030",
    doi = "10.1103/PhysRevD.92.074032",
    journal = "Phys. Rev. D",
    volume = "92",
    number = "7",
    pages = "074032",
    year = "2015"
}

@article{Chen:2016zka,
    author = "Chen, X. and Cruz-Martinez, J. and Gehrmann, T. and Glover, E. W. N. and Jaquier, M.",
    title = "{NNLO QCD corrections to Higgs boson production at large transverse momentum}",
    eprint = "1607.08817",
    archivePrefix = "arXiv",
    primaryClass = "hep-ph",
    reportNumber = "IPPP-16-73, ZU-TH-29-16, FR-PHENO-2016-014",
    doi = "10.1007/JHEP10(2016)066",
    journal = "JHEP",
    volume = "10",
    pages = "066",
    year = "2016"
}

@article{Chen:1977oja,
    author = "Chen, Kuo-Tsai",
    title = "{Iterated path integrals}",
    doi = "10.1090/S0002-9904-1977-14320-6",
    journal = "Bull. Am. Math. Soc.",
    volume = "83",
    pages = "831--879",
    year = "1977"
}

@article{Chicherin:2020oor,
    author = "Chicherin, Dmitry and Sotnikov, Vasily",
    title = "{Pentagon Functions for Scattering of Five Massless Particles}",
    eprint = "2009.07803",
    archivePrefix = "arXiv",
    primaryClass = "hep-ph",
    reportNumber = "MPP-2020-171",
    doi = "10.1007/JHEP12(2020)167",
    journal = "JHEP",
    volume = "20",
    pages = "167",
    year = "2020"
}

@article{Chicherin:2021dyp,
    author = "Chicherin, Dmitry and Sotnikov, Vasily and Zoia, Simone",
    title = "{Pentagon functions for one-mass planar scattering amplitudes}",
    eprint = "2110.10111",
    archivePrefix = "arXiv",
    primaryClass = "hep-ph",
    reportNumber = "LAPTH-041/21, MPP-2021-182",
    doi = "10.1007/JHEP01(2022)096",
    journal = "JHEP",
    volume = "01",
    pages = "096",
    year = "2022"
}

@article{Liu:2022chg,
    author = "Liu, Xiao and Ma, Yan-Qing",
    title = "{AMFlow: A Mathematica package for Feynman integrals computation via auxiliary mass flow}",
    eprint = "2201.11669",
    archivePrefix = "arXiv",
    primaryClass = "hep-ph",
    doi = "10.1016/j.cpc.2022.108565",
    journal = "Comput. Phys. Commun.",
    volume = "283",
    pages = "108565",
    year = "2023"
}

@article{Peraro:2019svx,
    author = "Peraro, Tiziano",
    title = "{$\text{FiniteFlow}$: multivariate functional reconstruction using finite fields and dataflow graphs}",
    eprint = "1905.08019",
    archivePrefix = "arXiv",
    primaryClass = "hep-ph",
    reportNumber = "ZU-TH 24/19",
    doi = "10.1007/JHEP07(2019)031",
    journal = "JHEP",
    volume = "07",
    pages = "031",
    year = "2019"
}

@article{vonManteuffel:2012np,
    author = "von Manteuffel, A. and Studerus, C.",
    title = "{Reduze 2 - Distributed Feynman Integral Reduction}",
    eprint = "1201.4330",
    archivePrefix = "arXiv",
    primaryClass = "hep-ph",
    reportNumber = "ZU-TH-01-12, MZ-TH-12-03, BI-TP-2012-02",
    month = "1",
    year = "2012"
}

@article{Henn:2020lye,
    author = "Henn, Johannes and Mistlberger, Bernhard and Smirnov, Vladimir A. and Wasser, Pascal",
    title = "{Constructing d-log integrands and computing master integrals for three-loop four-particle scattering}",
    eprint = "2002.09492",
    archivePrefix = "arXiv",
    primaryClass = "hep-ph",
    reportNumber = "MPP-2020-13, MITP/20-006, MIT-CTP/5176",
    doi = "10.1007/JHEP04(2020)167",
    journal = "JHEP",
    volume = "04",
    pages = "167",
    year = "2020"
}

@article{Baikov:1996iu,
    author = "Baikov, P. A.",
    editor = "Werlen, M. and Perret-Gallix, D.",
    title = "{Explicit solutions of the multiloop integral recurrence relations and its application}",
    eprint = "hep-ph/9611449",
    archivePrefix = "arXiv",
    reportNumber = "INP-96-42-449",
    doi = "10.1016/S0168-9002(97)00126-5",
    journal = "Nucl. Instrum. Meth. A",
    volume = "389",
    pages = "347--349",
    year = "1997"
}

@article{Frellesvig:2017aai,
    author = "Frellesvig, Hjalte and Papadopoulos, Costas G.",
    title = "{Cuts of Feynman Integrals in Baikov representation}",
    eprint = "1701.07356",
    archivePrefix = "arXiv",
    primaryClass = "hep-ph",
    doi = "10.1007/JHEP04(2017)083",
    journal = "JHEP",
    volume = "04",
    pages = "083",
    year = "2017"
}

@article{Gehrmann:2014bfa,
    author = "Gehrmann, Thomas and von Manteuffel, Andreas and Tancredi, Lorenzo and Weihs, Erich",
    title = "{The two-loop master integrals for $q\overline{q} \to VV$}",
    eprint = "1404.4853",
    archivePrefix = "arXiv",
    primaryClass = "hep-ph",
    reportNumber = "ZU-TH-16-14, LPN14-065-, MITP-14-021, LPN14-065",
    doi = "10.1007/JHEP06(2014)032",
    journal = "JHEP",
    volume = "06",
    pages = "032",
    year = "2014"
}

@article{Aliaj:2024zgp,
    author = "Aliaj, Rigers and Papathanasiou, Georgios",
    title = "{An Exceptional Cluster Algebra for Higgs plus Jet Production}",
    eprint = "2408.14544",
    archivePrefix = "arXiv",
    primaryClass = "hep-th",
    reportNumber = "DESY-24-129",
    month = "8",
    year = "2024"
}

@article{Cepeda:2019klc,
    author = "Cepeda, M. and others",
    editor = "Dainese, Andrea and Mangano, Michelangelo and Meyer, Andreas B. and Nisati, Aleandro and Salam, Gavin and Vesterinen, Mika Anton",
    title = "{Report from Working Group 2}: {Higgs Physics at the HL-LHC and HE-LHC}",
    eprint = "1902.00134",
    archivePrefix = "arXiv",
    primaryClass = "hep-ph",
    reportNumber = "CERN-LPCC-2018-04",
    doi = "10.23731/CYRM-2019-007.221",
    journal = "CERN Yellow Rep. Monogr.",
    volume = "7",
    pages = "221--584",
    year = "2019"
}

@article{Boughezal:2015dva,
    author = "Boughezal, Radja and Focke, Christfried and Liu, Xiaohui and Petriello, Frank",
    title = "{$W$-boson production in association with a jet at next-to-next-to-leading order in perturbative QCD}",
    eprint = "1504.02131",
    archivePrefix = "arXiv",
    primaryClass = "hep-ph",
    doi = "10.1103/PhysRevLett.115.062002",
    journal = "Phys. Rev. Lett.",
    volume = "115",
    number = "6",
    pages = "062002",
    year = "2015"
}

@article{Boughezal:2015ded,
    author = "Boughezal, Radja and Campbell, John M. and Ellis, R. Keith and Focke, Christfried and Giele, Walter T. and Liu, Xiaohui and Petriello, Frank",
    title = "{Z-boson production in association with a jet at next-to-next-to-leading order in perturbative QCD}",
    eprint = "1512.01291",
    archivePrefix = "arXiv",
    primaryClass = "hep-ph",
    reportNumber = "FERMILAB-PUB-15-519-T, IPPP-15-79",
    doi = "10.1103/PhysRevLett.116.152001",
    journal = "Phys. Rev. Lett.",
    volume = "116",
    number = "15",
    pages = "152001",
    year = "2016"
}

@article{Gehrmann:2024tds,
    author = "Gehrmann, Thomas and Henn, Johannes and Jakub{\v{c}}{\'\i}k, Petr and Lim, Jungwon and Mella, Cesare Carlo and Syrrakos, Nikolaos and Tancredi, Lorenzo and Torres Bobadilla, William J.",
    title = "{Graded transcendental functions: an application to four-point amplitudes with one off-shell leg}",
    eprint = "2410.19088",
    archivePrefix = "arXiv",
    primaryClass = "hep-th",
    reportNumber = "TUM-HEP 1529/24, MPP-2024-193, ZU-TH 52/24",
    doi = "10.1007/JHEP12(2024)215",
    journal = "JHEP",
    volume = "12",
    pages = "215",
    year = "2024"
}

@article{Chen:2025utl,
    author = "Chen, Xiang and Guan, Xin and Mistlberger, Bernhard",
    title = "{Three-loop QCD corrections to the production of a Higgs boson and a jet}",
    eprint = "2504.06490",
    archivePrefix = "arXiv",
    primaryClass = "hep-ph",
    reportNumber = "SLAC-PUB-250331, ZU-TH 24/25",
    doi = "10.1103/bqs9-2g2c",
    journal = "Phys. Rev. D",
    volume = "114",
    number = "1",
    pages = "L011502",
    year = "2026"
}

@article{Guan:2026igf,
    author = "Guan, Xin and Mistlberger, Bernhard and Ruf, Michael S.",
    title = "{Three-Loop QCD Corrections to Scattering Amplitudes of a Higgs Boson and Three Partons}",
    eprint = "2607.12033",
    archivePrefix = "arXiv",
    primaryClass = "hep-ph",
    reportNumber = "SLAC-PUB-260710",
    month = "7",
    year = "2026"
}

@article{Lee:2014ioa,
    author = "Lee, Roman N.",
    title = "{Reducing differential equations for multiloop master integrals}",
    eprint = "1411.0911",
    archivePrefix = "arXiv",
    primaryClass = "hep-ph",
    doi = "10.1007/JHEP04(2015)108",
    journal = "JHEP",
    volume = "04",
    pages = "108",
    year = "2015"
}

@article{Gituliar:2017vzm,
    author = "Gituliar, Oleksandr and Magerya, Vitaly",
    title = "{Fuchsia: a tool for reducing differential equations for Feynman master integrals to epsilon form}",
    eprint = "1701.04269",
    archivePrefix = "arXiv",
    primaryClass = "hep-ph",
    reportNumber = "DESY-16-219",
    doi = "10.1016/j.cpc.2017.05.004",
    journal = "Comput. Phys. Commun.",
    volume = "219",
    pages = "329--338",
    year = "2017"
}

@article{Arkani-Hamed:2010zjl,
    author = "Arkani-Hamed, Nima and Bourjaily, Jacob L. and Cachazo, Freddy and Caron-Huot, Simon and Trnka, Jaroslav",
    title = "{The All-Loop Integrand For Scattering Amplitudes in Planar N=4 SYM}",
    eprint = "1008.2958",
    archivePrefix = "arXiv",
    primaryClass = "hep-th",
    doi = "10.1007/JHEP01(2011)041",
    journal = "JHEP",
    volume = "01",
    pages = "041",
    year = "2011"
}

@article{Arkani-Hamed:2010pyv,
    author = "Arkani-Hamed, Nima and Bourjaily, Jacob L. and Cachazo, Freddy and Trnka, Jaroslav",
    title = "{Local Integrals for Planar Scattering Amplitudes}",
    eprint = "1012.6032",
    archivePrefix = "arXiv",
    primaryClass = "hep-th",
    doi = "10.1007/JHEP06(2012)125",
    journal = "JHEP",
    volume = "06",
    pages = "125",
    year = "2012"
}

@article{Bern:2014kca,
    author = "Bern, Zvi and Herrmann, Enrico and Litsey, Sean and Stankowicz, James and Trnka, Jaroslav",
    title = "{Logarithmic Singularities and Maximally Supersymmetric Amplitudes}",
    eprint = "1412.8584",
    archivePrefix = "arXiv",
    primaryClass = "hep-th",
    reportNumber = "CALT-TH-2014-172, UCLA-14-TEP-109",
    doi = "10.1007/JHEP06(2015)202",
    journal = "JHEP",
    volume = "06",
    pages = "202",
    year = "2015"
}

@article{Herrmann:2019upk,
    author = "Herrmann, Enrico and Parra-Martinez, Julio",
    title = "{Logarithmic forms and differential equations for Feynman integrals}",
    eprint = "1909.04777",
    archivePrefix = "arXiv",
    primaryClass = "hep-th",
    doi = "10.1007/JHEP02(2020)099",
    journal = "JHEP",
    volume = "02",
    pages = "099",
    year = "2020"
}

@article{Cachazo:2008vp,
    author = "Cachazo, Freddy",
    title = "{Sharpening The Leading Singularity}",
    eprint = "0803.1988",
    archivePrefix = "arXiv",
    primaryClass = "hep-th",
    month = "3",
    year = "2008"
}

@article{Guan:2025awp,
    author = "Guan, Xin and Mistlberger, Bernhard and Ruf, Michael",
    title = "{Subleading Color Corrections at Three Loops to the tr({\ensuremath{\phi}}$^2$) Three-Point Form Factor in N=4 Supersymmetric Yang-Mills Theory}",
    eprint = "2511.13845",
    archivePrefix = "arXiv",
    primaryClass = "hep-th",
    doi = "10.1103/hhhf-q3vb",
    journal = "Phys. Rev. Lett.",
    volume = "136",
    number = "16",
    pages = "161601",
    year = "2026"
}

@article{Gehrmann-DeRidder:2015wbt,
    author = "Gehrmann-De Ridder, A. and Gehrmann, T. and Glover, E. W. N. and Huss, A. and Morgan, T. A.",
    title = "{Precise QCD predictions for the production of a Z boson in association with a hadronic jet}",
    eprint = "1507.02850",
    archivePrefix = "arXiv",
    primaryClass = "hep-ph",
    reportNumber = "IPPP-15-44, ZU-TH-23-15",
    doi = "10.1103/PhysRevLett.117.022001",
    journal = "Phys. Rev. Lett.",
    volume = "117",
    number = "2",
    pages = "022001",
    year = "2016"
}

@article{Gehrmann-DeRidder:2017mvr,
    author = "Gehrmann-De Ridder, A. and Gehrmann, T. and Glover, E. W. N. and Huss, A. and Walker, D. M.",
    title = "{Next-to-Next-to-Leading-Order QCD Corrections to the Transverse Momentum Distribution of Weak Gauge Bosons}",
    eprint = "1712.07543",
    archivePrefix = "arXiv",
    primaryClass = "hep-ph",
    reportNumber = "CERN-TH-2017-280, IPPP-17-109, ZU-TH-38-17",
    doi = "10.1103/PhysRevLett.120.122001",
    journal = "Phys. Rev. Lett.",
    volume = "120",
    number = "12",
    pages = "122001",
    year = "2018"
}

@article{Alioli:2025hpa,
    author = "Alioli, Simone and Billis, Georgios and Broggio, Alessandro and Stagnitto, Giovanni",
    title = "{NNLO predictions with nonlocal subtractions and fiducial power corrections in GENEVA}",
    eprint = "2504.11357",
    archivePrefix = "arXiv",
    primaryClass = "hep-ph",
    reportNumber = "PSI-PR-25-08, UWThPh 2025-14",
    doi = "10.1007/JHEP01(2026)065",
    journal = "JHEP",
    volume = "01",
    pages = "065",
    year = "2026"
}

@article{NNLOJET:2025rno,
    author = "Huss, Alexander and others",
    collaboration = "NNLOJET",
    title = "{NNLOJET: A parton-level event generator for jet cross sections at NNLO QCD accuracy}",
    eprint = "2503.22804",
    archivePrefix = "arXiv",
    primaryClass = "hep-ph",
    reportNumber = "CERN-TH-2025-012, IPPP/25/09, ZU-TH 11/25",
    doi = "10.21468/SciPostPhysCodeb.69",
    journal = "SciPost Phys. Codeb.",
    volume = "69",
    pages = "1",
    year = "2026"
}

@article{Garland:2002ak,
    author = "Garland, L. W. and Gehrmann, T. and Glover, E. W. Nigel and Koukoutsakis, A. and Remiddi, E.",
    title = "{Two loop QCD helicity amplitudes for $e^+e^-\to$ three jets}",
    eprint = "hep-ph/0206067",
    archivePrefix = "arXiv",
    reportNumber = "IPPP-02-28, DCPT-02-56, CERN-TH-2002-122",
    doi = "10.1016/S0550-3213(02)00627-2",
    journal = "Nucl. Phys. B",
    volume = "642",
    pages = "227--262",
    year = "2002"
}

@article{Magerya:2022hvj,
    author = "Magerya, Vitaly",
    title = "{Rational Tracer: a Tool for Faster Rational Function Reconstruction}",
    eprint = "2211.03572",
    archivePrefix = "arXiv",
    primaryClass = "physics.data-an",
    reportNumber = "KA-TP-26-2022, P3H-22-109",
    month = "11",
    year = "2022"
}

@article{Henn:2019swt,
    author = "Henn, Johannes M. and Korchemsky, Gregory P. and Mistlberger, Bernhard",
    title = "{The full four-loop cusp anomalous dimension in $\mathcal{N}=4$ super Yang-Mills and QCD}",
    eprint = "1911.10174",
    archivePrefix = "arXiv",
    primaryClass = "hep-th",
    reportNumber = "IPhT-T19/155, MIT-CTP/5160, MPP-2019-233",
    doi = "10.1007/JHEP04(2020)018",
    journal = "JHEP",
    volume = "04",
    pages = "018",
    year = "2020"
}

@article{Czakon:2026tbd,
    author = "Czakon, Micha{\l} and Tancredi, Lorenzo",
    title = "{Solution of Canonical Differential Equations for Integrals on Arbitrary Geometries}",
    eprint = "2606.30354",
    archivePrefix = "arXiv",
    primaryClass = "hep-ph",
    year = "2026"
}

@article{Ahnert:2011odeint,
    author = "Ahnert, Karsten and Mulansky, Mario",
    title = "{Odeint -- Solving Ordinary Differential Equations in C++}",
    eprint = "1110.3397",
    archivePrefix = "arXiv",
    journal = "AIP Conf. Proc.",
    volume = "1389",
    pages = "1586",
    year = "2011"
}

\end{document}